\documentclass[final,5p,times,twocolumn,authoryear]{elsarticle}

\usepackage{epsfig}
\usepackage{verbatim} 
\usepackage{epstopdf} 
\usepackage{graphicx}
\usepackage{aas_macros}
\usepackage{epsf}
\usepackage{amsmath}
\usepackage{xcolor}
\usepackage{tikz}
\usepackage{longtable}
\usepackage{deluxetable}

\definecolor{darkgreen}{rgb}{0.0,0.5,0.0}
\usepackage[colorlinks,citecolor=darkgreen]{hyperref} 

\newcommand{\ie}{\emph{i.e.} }
\newcommand{\eg}{\emph{e.g.,} }

\newcommand{\be}{\begin{equation}}
\newcommand{\ee}{\end{equation}}
\newcommand{\bea}{\begin{equation*}}
\newcommand{\eea}{\end{equation*}}
\newcommand{\beqr}{\begin{eqnarray} \nonumber}
\newcommand{\eeqr}{\end{eqnarray}}
\newcommand{\beqrb}{\begin{eqnarray}}
\newcommand{\eeqrb}{\nonumber \end{eqnarray}}
\newcommand{\fin}{\mbox{ .}}
\newcommand{\coma}{\mbox{ ,}}

\newcommand{\cm}{\mbox{ cm}}
\newcommand{\sr}{\mbox{ sr}}
\newcommand{\se}{\mbox{ s}}
\newcommand{\yr}{\mbox{ yr}}

\newcommand{\erg}{\mbox{ erg}}

\newcommand{\km}{\mbox{ km}}

\newcommand{\Mpc}{\mbox{ Mpc}}

\newcommand{\GeV}{\mbox{ GeV}}

\newcommand{\gama}{$\gamma$}

\newcommand{\mynewcommand}[2]{\ifdefined #1 \else \newcommand{#1}{#2} \fi}

\mynewcommand{\apj}{ApJ}     
\mynewcommand{\apjl}{ApJL}     
\mynewcommand{\apjs}{ApJS}    
\mynewcommand{\aap}{A\&A}    
\mynewcommand{\nat}{Nature}  

\newcommand{\dgr}{^{\circ}}

\newcommand{\dgrdot}{{\overset{^\circ}{.}}}

\newcommand{\bmt}{{\boldsymbol{\theta}}}
\newcommand{\bmr}{{\boldsymbol{r}}}
\newcommand{\bmp}{{\boldsymbol{\psi}}}
\newcommand{\myR}{R_{500}}
\newcommand{\Myfr}{{j}}
\newcommand{\Myc}{{\mathsf{c}}}
\newcommand{\MyC}{{\mathsf{C}}}
\newcommand{\Myh}{{\mathsf{h}}}
\newcommand{\MyH}{{\mathsf{H}}}
\newcommand{\mys}{{s}}
\newcommand{\Myn}{{\mathsf{n}}}
\newcommand{\Myf}{{\mathsf{f}}}
\newcommand{\Fukazawa}{F04}
\newcommand{\Chen}{C07}

\defcitealias{ReissKeshet18}{Paper I}
\newcommand{\PapI}{{\citetalias{ReissKeshet18}}}

    \makeatletter
\def\ps@pprintTitle{%
  \let\@oddhead\@empty
  \let\@evenhead\@empty
  \def\@oddfoot{\reset@font\hfil\thepage\hfil}
  \let\@evenfoot\@oddfoot
}
\makeatother

\begin{document}

\begin{frontmatter}

\title{Galaxy-cluster-stacked Fermi-LAT II: extended central hadronic signal}

\author{Uri Keshet}
\address{
    Physics Department, Ben-Gurion University of the Negev, POB 653, Be'er-Sheva 84105, Israel; keshet.uri@gmail.com
}

\begin{abstract}
Faint $\gamma$-ray signatures emerge in \emph{Fermi}-LAT data stacked scaled to the characteristic $\theta_{500}$ angles of MCXC galaxy clusters.
After Paper I of this series thus discovered virial shocks, later supported in other bands, this second paper focuses on cluster cores.
Stacking $1$--$100$ GeV source-masked data around clusters shows a significant ($4.7\sigma$ for 75 clusters)
and extended central excess, inconsistent with central point sources.
The resolved signal is best fit ($3.7\sigma$ TS-test) as hadronic emission from a cosmic-ray ion (CRI) distribution that is flat both spectrally ($p\equiv1-d\ln u/d\ln E=2.0\pm0.3$) and spatially (CRI-to-gas index $\sigma=0.1\pm0.4$), carrying an energy density $du(0.1\theta_{500})/d\ln E=10^{-13.6\pm0.5}$ erg cm$^{-3}$ at $E=100$ GeV energy; insufficient resolution would raise $p$ and $\sigma$.
Such CRI match the long-predicted distribution needed to power diffuse intracluster radio emission in its various forms (mini-halos, giant halos, standard relics, their transitional forms, and mega-halos), disfavoring models that invoke electron (re)acceleration in weak shocks or turbulence.
Stringent upper limits on residual $\gamma$-ray emission, \emph{e.g.} from dark-matter annihilation, are imposed.
\end{abstract}

\end{frontmatter}

\section{Introduction}

Cosmic-ray (CR) ions and electrons in the intracluster medium (ICM) of clusters or groups of galaxies (clusters, henceforth) have subtle radiative signatures that can be used to map the CR distribution, trace related physical processes, and elucidate the CR origin, dynamical role, and impact on ICM evolution.
High-energy CR electrons (CRE) cool rapidly, so their radiative signature traces recent particle acceleration in strong virial and supernova-remnant shocks; weak merger shocks were not proven to provide efficient CRE acceleration.
CR ions (CRI) cool slowly, so they accumulate through advection and diffusion in the ICM, tracing its dense regions where inelastic collisions with ambient nuclei become more frequent.

\begin{figure}
    \centering
    \includegraphics[width=0.45\textwidth]{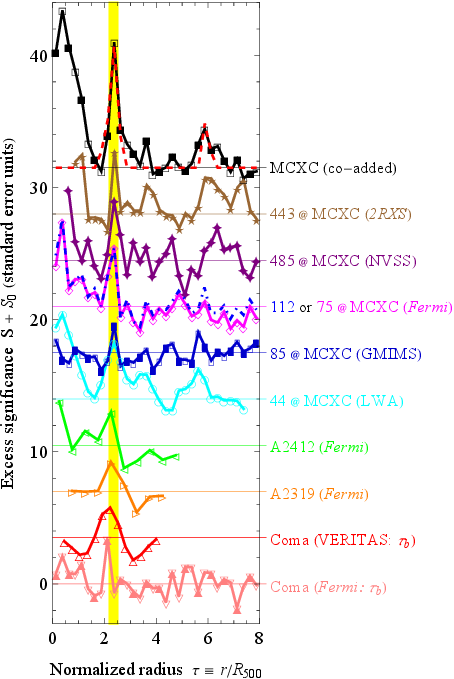}
	\caption{
        The \emph{Fermi}-LAT signal of {\PapI}, stacked over 112 clusters (dot-dashed blue curve), is updated (nominal analysis: magenta diamonds), showing equally significant central and virial-shock signals with the same data using only 75 clusters, thanks to the improved 4FGL-DR4 catalog.
        Each signal is presented as the significance $S$ (symbols with lines to guide the eye, in standard-error units) of the excess above the background $S_0$ (labelled horizontal lines, shifted vertically for visibility) as a function of the normalized radius $\tau$.
        Follow-up MCXC stacking analyses (bottom to top labels specify sample sizes) have shown additional signals well-localized in the same region ($2.2<\tau<2.5$; vertical yellow band), in diffuse emission (empty symbols), discrete sources (filled), or without separating the two (intermittent empty and filled symbols): in LWA \cite[circles;][]{HouEtAl23} and polarized GMIMS \cite[rectangles;][]{Keshet24GMIMS}, and in NVSS (four-pointed stars) and 2XRS (five-stars) source catalogs \citep{IlaniEtAl24a}; the co-added MCXC excess (black squares) is consistent with a stacked cylindrical shock model \cite[dashed red curve;][]{Keshet24GMIMS}.
        Also shown (triangles) are virial shock signals in individual clusters (bottom to top) Coma, in LAT \cite[down triangles;][]{keshet2018evidence} and VERITAS \cite[virial excess coincident with synchrotron emission; up triangles;][]{KeshetEtAl17} data as a function of $\tau_b$, and A2319 (right triangles) and A2142 (left triangles) in LAT data \citep{keshet20coincident}, all coincident with a drop in Sunyaev-Zel'dovich (SZ) signal.
    \label{fig:VSummary}
	}
\end{figure}

Such faint signals can be amplified by stacking data over many clusters, each normalized by a characteristic length scale such as the $R_{500}$ radius, which encloses a mean density $500$ times the critical mass density of the Universe.
Paper I \citep{ReissKeshet18} thus stacked \emph{Fermi}-LAT data over 112 high-latitude clusters from the Meta-Catalog of X-ray detected Clusters of galaxies \cite[MCXC;][]{PiffarettiEtAl11}, finding a significant excess at a normalized radius $\tau\equiv r/R_{500}\simeq 2.3\pm 0.1$.
This signal was identified as inverse-Compton emission from CRE accelerated to a
flat, index $p\equiv -d\ln N/d\ln E =2.1\pm0.2$ energy $E$ spectrum by the strong cluster virial shock, consistent with signals detected earlier in the Coma cluster \citep{KeshetEtAl17, keshet2018evidence}.
Several other highly localized $\tau\simeq 2.3$ signals were subsequently discovered, as illustrated in Fig.~\ref{fig:VSummary}, confirming the virial shock detection and constraining the physics of such non-magnetized collisionless shocks \citep{KeshetHou24}.

The strongest CRI signature is expected from the core of the cluster, where the high ICM density implies a maximal rate of inelastic pion-producing collisions leading primarily to $\pi^0\to\gamma\gamma$ emission.
Several \emph{Fermi}-LAT stacking analyses that did not normalize scales by $R_{500}$ could not find such a \gama-ray signal \citep[\eg][]{HuberEtAl13, AckermannEtAl14_GammaRayLimits, ProkhorovChurazov14, GriffinEtAl14}, imposing only upper limits on the central flux.
A central \gama-ray signal in Coma was recently reported \citep{XiEtAl18_Coma, AdamEtAl21, BaghmanyanEtAl22}, but the results differ among studies for such a single, extended cluster, as they are sensitive to modeling and details such as cluster morphology, foreground gradients, nearby point sources that may contaminate the entire core, and the assumed spectrospatial CRI distribution \citep{Keshet24, KushnirEtAl24}.

An alternative CRI tracer is radio synchrotron emission of secondary CRE produced from the decay of $\pi^\pm$ produced in the same inelastic $p$-$p$ collisions.
Such a mechanism was proposed for giant halos \citep[GH;][]{Dennison80}, minihalos \citep[MH;][]{KeshetLoeb10}, standard relics \citep{Keshet10}, and mega-halos \citep{Keshet24}. 
Observations such as MH--GH transitions and GH--relic bridges support a joint hadronic model underlying these different sources \citep{Keshet10, Keshet24}.
Indeed, diffuse ICM radio emission in its various forms is naturally explained by the same extended CRI population, best fit by a homogeneous, spectrally-flat ($p\simeq 2$) CRI distribution of differential energy density \citep{Keshet10}
\begin{equation}\label{eq:CRI_prediction}
 \frac{du}{d\ln E} = 10^{-13.5\pm0.5}\erg\cm^{-3}\fin
\end{equation}
Additionally, the spectrospatial properties of individual sources imply strong CR diffusion or mixing, sufficient for CRI homogenization \citep{Keshet24}.

It should be mentioned that such hadronic (secondary CRE) halo models are increasingly dismissed in recent years \citep[\eg][and references therein]{vanWeerenEtAl19} in favor of primary CRE models, typically invoking CRE (re)acceleration in turbulence attributed to mergers in GHs \citep{EnsslinEtAl99} and to sloshing in MHs \citep{MazzottaGiacintucci08}, while relics are usually modeled as powered by diffusive CRE acceleration in weak merger shocks \citep{EnsslinEtAl98}; for a discussion of the difficulties and inconsistencies of such models, see \citet{Keshet10, Keshet24}.
The dismissal of hadronic models is often blamed on the absence of a \gama-ray counterpart.
For instance, when the signal was detected in Coma, it was claimed to be too weak, by a factor of a few, than needed to match the GH \citep{AdamEtAl21}.
However, systematic errors are larger than a factor of a few; indeed, \gama-rays and radio were later shown to be in good agreement in Coma if one accounts for the extended and hard CRI distribution \citep{Keshet24, KushnirEtAl24}.

Here, we update and resume the scaled MCXC stacking analysis initiated in {\PapI}, now using the updated \emph{Fermi}-LAT Fourth Source Catalog 4FGL \citep[Data Release 4;][]{FGL4DR4} instead of the 3FGL catalog \citep{FermiPSC} utilized previously.
Avoiding clusters with nearby point sources from this new and improved catalog selects a somewhat different cluster sample. 
Therefore, we reproduce the virial-shock signal of {\PapI} (see Fig.~\ref{fig:VSummary}) before considering the central cluster regions, where we find and study a significant signal emerging at $r<0.5R_{500}$ radii.
The data and analysis methods are reviewed in \S\ref{sec:Methods}, and the properties of the central excess and other stacked signals are presented in \S\ref{sec:CentralExcess}. Different models are outlined in \S\ref{sec:Models}, and applied to the data in \S\ref{sec:Modelling}.
The results are finally summarized and discussed in \S\ref{sec:Discussion}.
Also presented are sensitivity tests (\ref{sec:Sensitivity}), empirical PSF estimates (\ref{sec:EmpiricalPSF}), tests of different model variants (\ref{sec:Variants}), and a list of clusters in our sample (\ref{sec:Sample}), along with their properties.

We adopt the same flat $\Lambda$CDM cosmological model used in MCXC, with a present-day Hubble constant $H_0=70\km\se^{-1}\Mpc^{-1}$ and a mass fraction $\Omega_m=0.3$; an $f_b=0.17$ baryon fraction is assumed.
Error bars indicate $68.3\%$ containment projected onto one parameter, whereas model parameter uncertainty values and contours reflect multivariate best-fit standard errors ($68.3\%$, $95.4\%$, $99.73\%$).

\section{Methods: data, catalogs, stacking, and binning}
\label{sec:Methods}

In order to analyze the central emission from clusters, we update the analysis of {\PapI}, adopting similar methods and selection criteria, but with up-to-date data and pipeline.
In particular, we focus on the same $\epsilon=1\mbox{--}100\GeV$ photon energy range (limited from below by LAT resolution and from above by photon statistics), and stack the data around $0.2<\theta_{500}<0.5$ (limited from below by resolution and from above by foreground structure and point-source abundance) MCXC clusters, avoiding in addition the Galactic plane and nearby point sources.
The main difference with respect to {\PapI} is the updated 4FGL-DR4 point-source catalog, which is based on the extended time period August 4, 2008, to August 2, 2022.

Namely, we use updated archival, Pass-8 \citep[P8R3;][]{BruelEtAl18Pass8R3} LAT data from the Fermi Science Support Center (FSSC)\footnote[1]{\texttt{http://fermi.gsfc.nasa.gov/ssc}}, analyzed with updated Fermi Science Tools (version \texttt{2.2.0}).
Pre-generated weekly all-sky files were binned using the corresponding P8R3\_ULTRACLEANVETO\_V3 instrument response functions (IRFs) into $N_\epsilon=4$ logarithmically-spaced energy bands in the (1--100) GeV range (labelled below as channels $j=1$ through 4).
A zenith angle cut of $90\dgr$ was applied to reduce atmospheric interference, and good time intervals were identified using the recommended selection expression \texttt{(DATA\_QUAL==1) and (LAT\_CONFIG==1)}.
Sky maps were discretized using a HEALPix scheme \citep{GorskiEtAl05} of order $N_{hp}=10$, providing a $\delta\Omega\simeq 10^{-6}\sr$ pixel solid angle and a mean $\sim 0\dgrdot057$ pixel separation, shown in {\PapI} to suffice for analyzing clusters above the $\theta_{500}>0\dgrdot2$ limit imposed by the point spread function \citep[PSF;][]{AtwoodEtAl13}.

In order to obtain a clean cluster sample and avoid spurious signals, we exclude clusters lying at low, $|b|<30\dgr$ latitudes, as well as all clusters found in the two over-dense MCXC regions identified in {\PapI}: the 12 clusters near Galactic coordinates $(l,b)=(315\dgr,32\dgr)$, and the 4 clusters near $(12\dgr,50\dgr)$.
As in {\PapI}, we minimize contamination from point sources by masking all HEALPix pixels found within the $95\%$ containment area (in each energy band) of any of the 5697 significant ($>5\sigma$) 4FGL-DR4 sources.

In addition, as in {\PapI}, we wish to exclude all clusters which have a point source within $1\dgrdot8$ (the $95\%$ containment angle at $1\GeV$) from their center, for extra caution.
However, 4FGL-DR4 contains more than twice the sources of 3FGL (7195 vs. 3034), many of which are identified at low significance or are poorly localized, so excluding clusters around all catalog sources would be excessive.
Therefore, for the purpose of cluster exclusion, we consider only the 3578 sources that are both significant ($>5\sigma$) and well localized ($95\%$ containment semi-major axis $a_{95}<0\dgrdot08$).
The $a_{95}$ cut is chosen larger than all 4FGL-DR4 sources lying in the centers of clusters, to guarantee that these clusters are all excluded from our sample.
As shown below, this $a_{95}$ scale is $>3$ times larger than the typical cluster core in our sample, so such sources may well be ICM emission not yet identified as extended.
We note in passing that none of the 82 sources identified as extended in 4FGL-DR4 lie in the vicinity of our sample clusters.

\begin{figure}[h!]
    \centering
    \includegraphics[trim={0 0.3cm 0 0}, clip, width=0.45\textwidth]{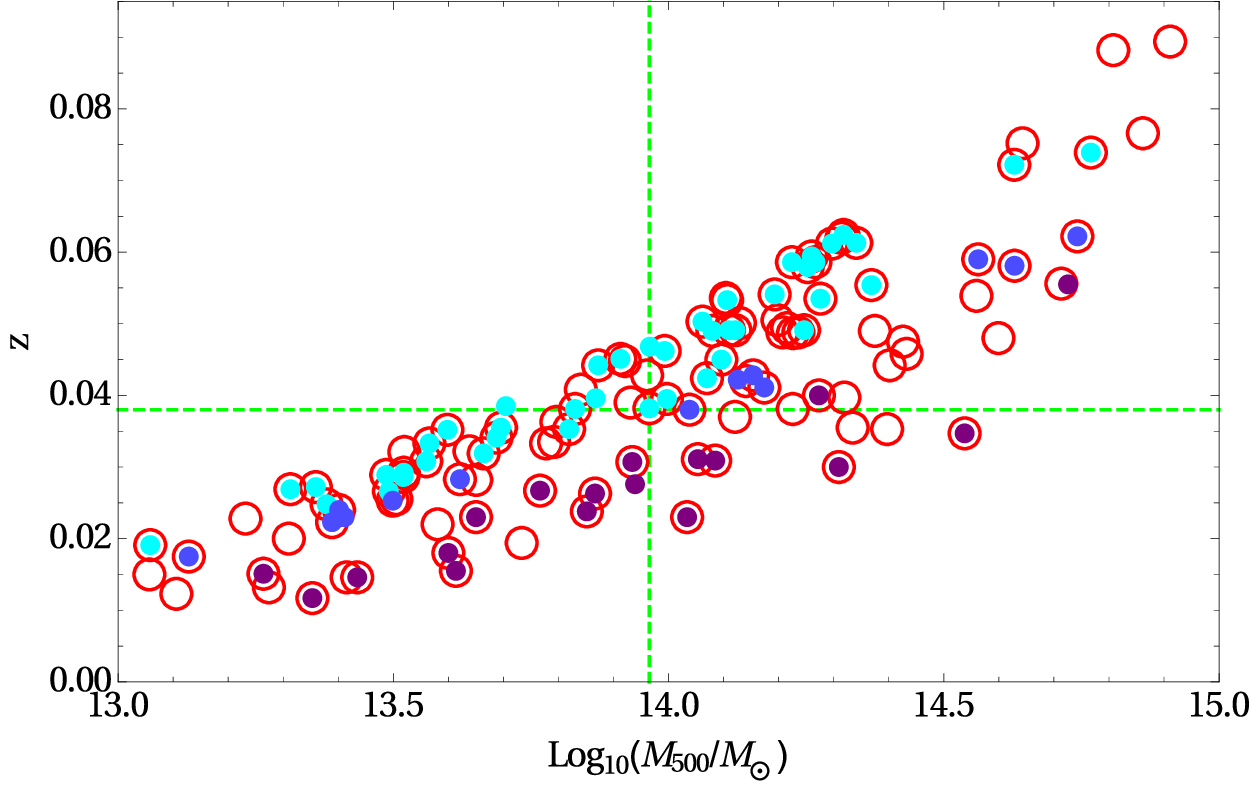}
   	\caption{
        Mass and redshift in our nominal sample of 75 high-latitude clusters (disks, with median values given by green dashed lines, colored light cyan for $\theta_{500}<0\dgrdot25$ and dark purple for $\theta_{500}>0\dgrdot3$) and of the {\PapI} sample of 112 clusters (red circles). High-redshift clusters are absent due to the $\theta_{500}>0\dgrdot2$ cut. The present sample is smaller than in {\PapI} due to the updated point-source catalog (4FGL DR4 instead of 3FGL), the more stringent latitude cut ($|b|>30\dgr$), and avoiding the two cluster-overlap regions pointed out in {\PapI}.
    }
    \label{fig:Sample75}
\end{figure}

The above considerations yield a nominal sample of $75$ clusters, listed in \ref{sec:Sample}.
Figure \ref{fig:Sample75} shows the clusters of this sample (disks), as well as the 112 clusters of {\PapI} (red circles), in the phase space of $M_{500}$ mass vs. redshift $z$.
The nominal sample has $\mbox{med}(M_{500})\simeq 9.2\times 10^{13}M_\odot$ mass and $\mbox{med}(z)\simeq 0.038$ redshift median values.
The overall results shown below are not sensitive to reasonable changes around our nominal cuts, but modeling the extent of the signal is resolution-limited, so we later consider also partial samples excluding the more compact $\theta_{500}<0\dgrdot25$ (cyan disks) clusters or including only the most extended $\theta_{500}>0\dgrdot3$ (purple disks) clusters of the nominal sample.

We analyze data from the extended period spanning weeks $9$ through $859$ ($16.3\yr $) of the \emph{Fermi} mission, but also consider two separate time periods: epoch I, examined in {\PapI}, spanning the earlier weeks $9$--$422$ ($7.9\yr$) of the mission, and epoch II, spanning the later weeks $423$--$859$ ($8.4\yr$).
We also subdivide each of these two epochs into four $\sim 100$-week periods, to test for possible changes in LAT performance.
While both the virial-shock signal of {\PapI} and the central emission analyzed here present significantly in the full mission data, we find that the signals were stronger in epoch I.
The virial-shock signal, in particular, shows independently in each of the four periods of epoch I, but in none of the four periods of epoch II.
As shown in \ref{sec:EmpiricalPSF}, stacking data around bright compact sources suggests that the PSF has degraded somewhat after epoch I, with an adverse effect on resolution-limited stacking analyses.
Our nominal analysis is thus based on the same epoch I studied in {\PapI}, although epoch I+II results are also examined.

We stack data in the photon co-addition method of {\PapI}.
Namely, for each energy channel $\Myfr$, cluster $\Myc$, and non-masked HEALPix pixel $\Myh$, let $\Delta \Myn_{\Myfr}(\Myc,\Myh) = \Myn_{\Myfr}(\Myh)-\Myf_{\Myfr}(\Myc,\Myh)$ be the excess in the number $\Myn$ of detected photons with respect to the expected number $\Myf$ of foreground and background (field, for simplicity) photons.
The mean excess brightness per cluster for a sample $\MyC$ of $N_c$ clusters then becomes
\begin{equation}
\label{eq:ExcessIj}
I_{\Myfr}(\tau; C; \Myf)
= \frac{\epsilon_{\Myfr}\sum_{\Myc\in\MyC} \sum_{\Myh\in \MyH} \Delta \Myn_{\Myfr}(\Myc,\Myh)}{N_c \delta\Omega \sum_{\Myc\in\MyC} \sum_{\Myh\in \MyH} \mathcal{E}_{\Myfr}(\Myh) } \coma
\end{equation}
where $\MyH(\tau,\Myc)$ are the non-masked HEALPix pixels falling in the $\tau$ bin of cluster $\Myc$, and $\mathcal{E}_{\Myfr}(\Myh)$ is the channel-$\Myfr$ exposure (\ie time-integrated effective area) of pixel $\Myh$.
The averaged photon energy $\epsilon_{\Myfr}$ in channel $\Myfr$ is computed, for simplicity, for a flat photon spectrum, $\mys\equiv 1-d\ln I_\epsilon/d\ln \epsilon=2$, where $I_\epsilon$ is the specific brightness (such that $I_j=\int_{\raisebox{0.3ex}{\scriptsize$j$}} I_\epsilon\,d\epsilon$);
this choice has a negligible effect on the results because our energy bins are narrow.

The field distribution $\Myf$ is evaluated around each cluster $\Myc$ by fitting the sky-distribution of the brightness $I_{\Myfr}(\tau_x,\tau_y; \{\Myc\};0)$ within the cluster region of interest (defined nominally as $\tau<15$) as a polynomial of order $N_f$ in the normalized sky coordinates $(\tau_x, \tau_y)$.
We choose $N_f=2$ as our nominal polynomial order, but also consider $N_f=0$ and $N_f=4$ orders (odd orders to not contribute to radial binning).

The significance of an $I_{\Myfr}>0$ excess can be estimated for $\Myf\gg1$ in the normal-distribution limit as
\begin{equation}
\label{eq:ExcessSj}
S_{\Myfr}(\tau; C; \Myf)
= \frac{\sum_{\Myc\in\MyC} \sum_{\Myh\in \MyH} \Delta \Myn_{\Myfr}}{\sqrt{\sum_{\Myc\in\MyC} \sum_{\Myh\in \MyH} \Myf_{\Myfr}} } \coma
\end{equation}
in standard-error units.
When combining $N_\epsilon$ channels, the excess brightness averaged over channels becomes
\begin{equation}
\label{eq:ExcessI}
\langle I_{\Myfr}\rangle(\tau; C; \Myf)
= \frac{1}{N_\epsilon} \sum_{j=1}^{N_\epsilon} I_{\Myfr} \coma
\end{equation}
corresponding to a co-added excess
\begin{equation}
\label{eq:ExcessS}
S(\tau; C; \Myf)
= \frac{1}{\sqrt{N_\epsilon}} \sum_{j=1}^{N_\epsilon} S_{\Myfr} \fin
\end{equation}
The results \eqref{eq:ExcessSj} and \eqref{eq:ExcessS} were tested for the real data using cluster control-samples in {\PapI}, and shown to hold well throughout the region of interest, and even for small $\tau$ where radial bins cover a small solid angle.
{\PapI} also showed, through extensive sensitivity tests, that the results are robust to reasonable changes in all selection cuts and analysis parameters; also see \ref{sec:Sensitivity}.

\section{Results: central and peripheral excess signals}
\label{sec:CentralExcess}

\begin{figure}[h!]
    \centering
    \includegraphics[width=0.457\textwidth]{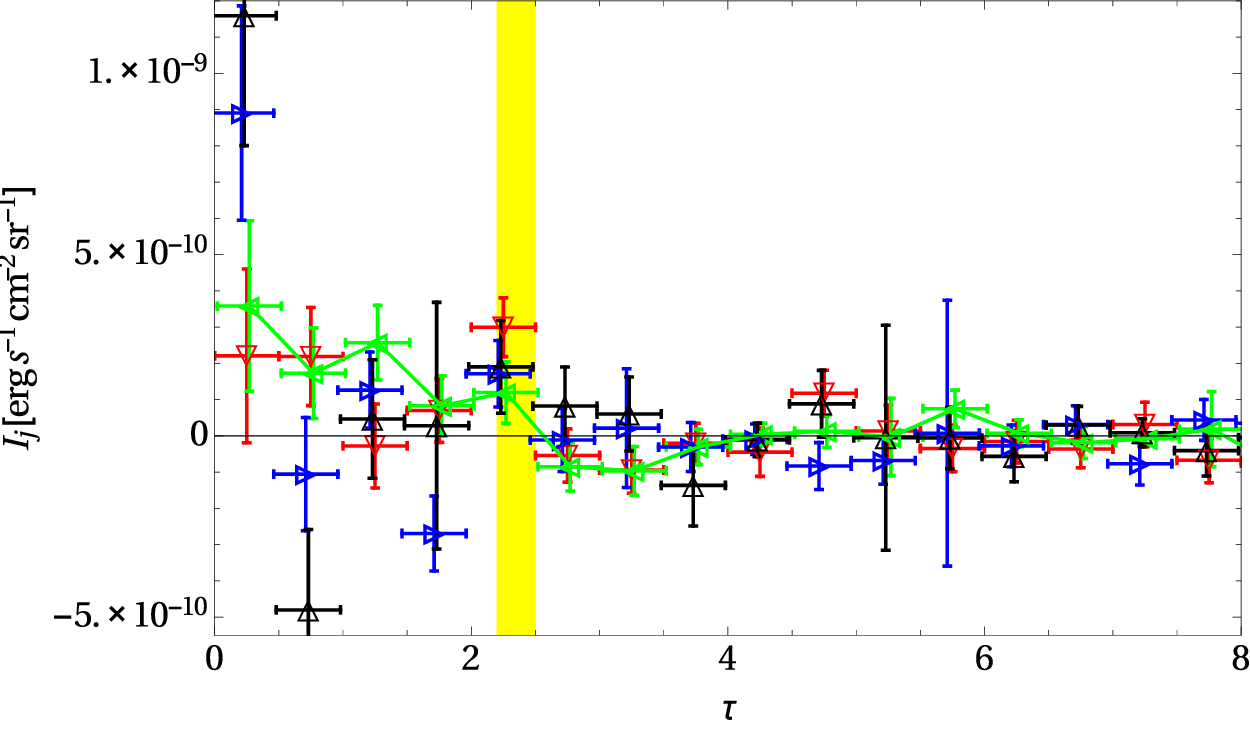}\\
    \hspace{0.2cm}\includegraphics[width=0.45\textwidth]{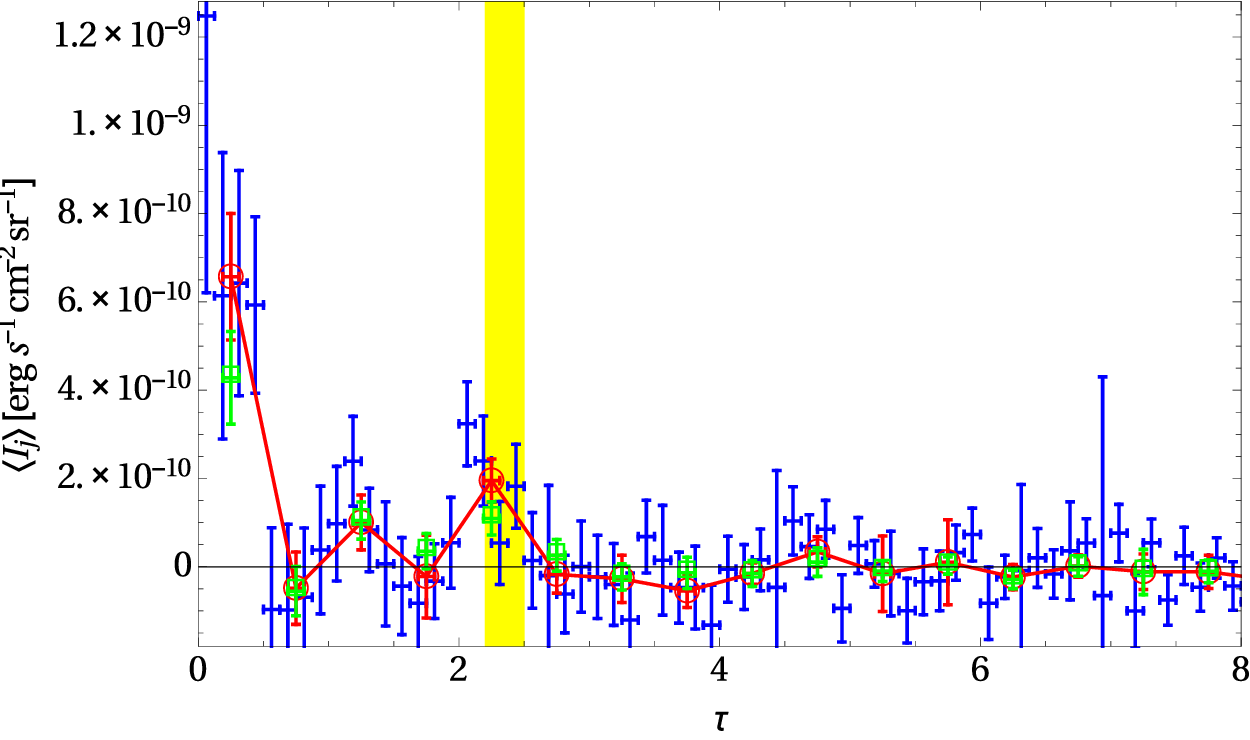}
   	\caption{
        Radially-binned brightness stacked for nominal parameters and sample: 75 high-latitude, $|b|>30\dgr$ clusters, using a second-order fitted foreground based on $\tau<15$ data.\\
        Top panel: separate energy channels 1 (red down-triangles), 2 (green left-triangles, with illustrative solid line to guide the eye), 3 (blue right-triangles), and 4 (black up-triangles) shown at low ($\Delta\tau=1/2$) resolution (top; error bars slightly shifted horizontally for visibility).\\
        Bottom panel: averages over the four channels shown at both low (red circles with error bars and line to guide the eye) and high ($\Delta\tau=1/8$; blue error bars) resolutions. Low resolution results also shown (greed squares with error bars) for epochs I+II. The virial-shock region is highlighted as in Fig.~\ref{fig:VSummary}.
        The central ($\tau<0.5$) emission is detected at a nominal $4.6\sigma$ confidence level.
    }
    \label{fig:Flux}
\end{figure}

Figure \ref{fig:Flux} shows the stacked brightness in our nominal analysis, radially binned at $\Delta\tau=1/2$ resolution, for each of the four channels (top panel) according to Eq.~\eqref{eq:ExcessIj}, and for the channel-averaged brightness (bottom panel) according to Eq.~\eqref{eq:ExcessI}; the corresponding significance of the excess is shown in Fig.~\ref{fig:VSummary} (magenta diamonds) according to Eq.~\eqref{eq:ExcessS}.
These figures indicate three distinct signals: (i) the strong, $4.7\sigma$ central excess in the $\tau<0.5$ core is the focus of the present paper; (ii) the $4.2\sigma$ virial-shock $2.0<\tau<2.5$ excess was the topic of {\PapI}; and (iii) a tentative $2.4\sigma$ intermediate, $1.0<\tau<1.5$ signal is deferred to Paper III of this series.
Using smaller, $\Delta\tau=1/8$ bins (in the bottom panel of Fig.~\ref{fig:Flux}) resolves all three signals in the two high-energy bins, where the $68\%$ containment radius is $<0\dgrdot1$, but not in the low energy bin where it exceeds $0\dgrdot3$.
Also presented (in the bottom panel) is the mean brightness for the full epoch I+II data, demonstrating that while these three signals persist over the full mission, they are dominated by epoch I, apparently due to subsequent PSF degradation (see \S\ref{sec:Methods} and \ref{sec:EmpiricalPSF}).

Figure \ref{fig:VSummary} shows that the virial-shock excess in the present analysis (magenta diamonds) is nearly identical to that of {\PapI} (dot-dashed blue curve), despite being stacked over only 75 clusters, as compared to the 112 clusters of {\PapI}, and even though point-source masking and cluster exclusion are now based on many more sources, taken from a better and larger catalog.
This $\tau\simeq 2.4$ excess, as well as other coincident virial-shock signals demonstrated in the figure, are better fit by a cylindrical shock model \citep[of $\tau\simeq 2.4$ radius, projected and stacked; dashed red curve; see][]{Keshet24GMIMS} than by a projected spherical shell.
In such a model, the stacked signal rapidly diminishes at $\tau\lesssim2$, explaining why locally it matches a ring on the sky \citep[the 'planar model' of {\PapI} and ][]{keshet2018evidence} better than a projected sphere.
The model also implies a secondary, external, broad excess, which is tentatively seen in the figure around $\tau\simeq 6$ (broad $3.7\sigma$ excess in the co-added, top black curve).
Such a virial-shock signal does not extended to small radii, so it can be neglected in the central-signal analysis below.

\begin{figure*}[t!]
    \centering
    \includegraphics[width=0.33\textwidth]{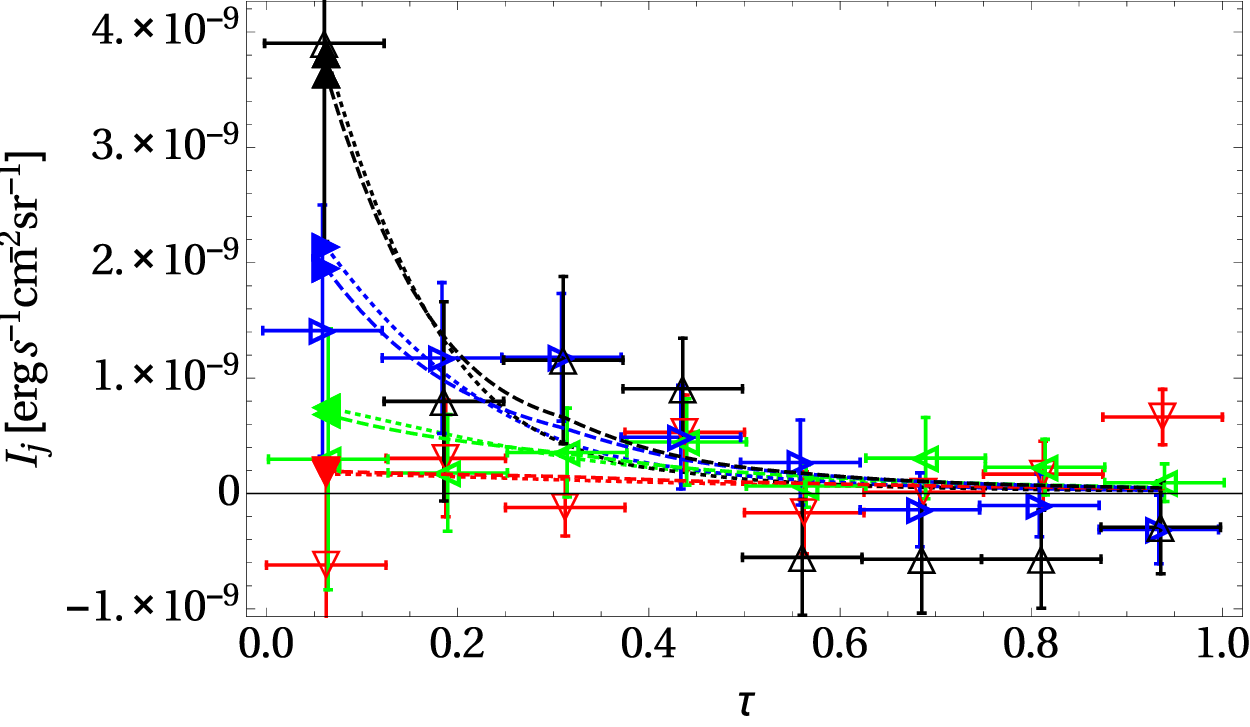}
    \includegraphics[width=0.33\textwidth]{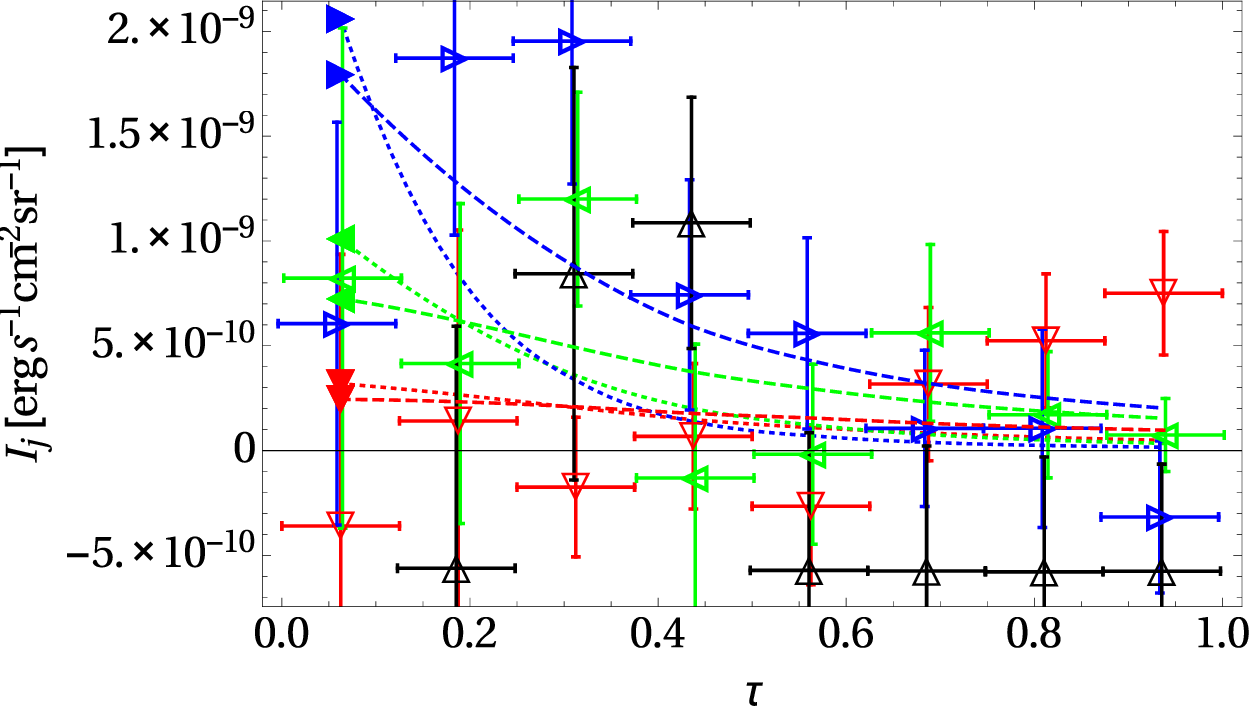}
    \includegraphics[width=0.33\textwidth]{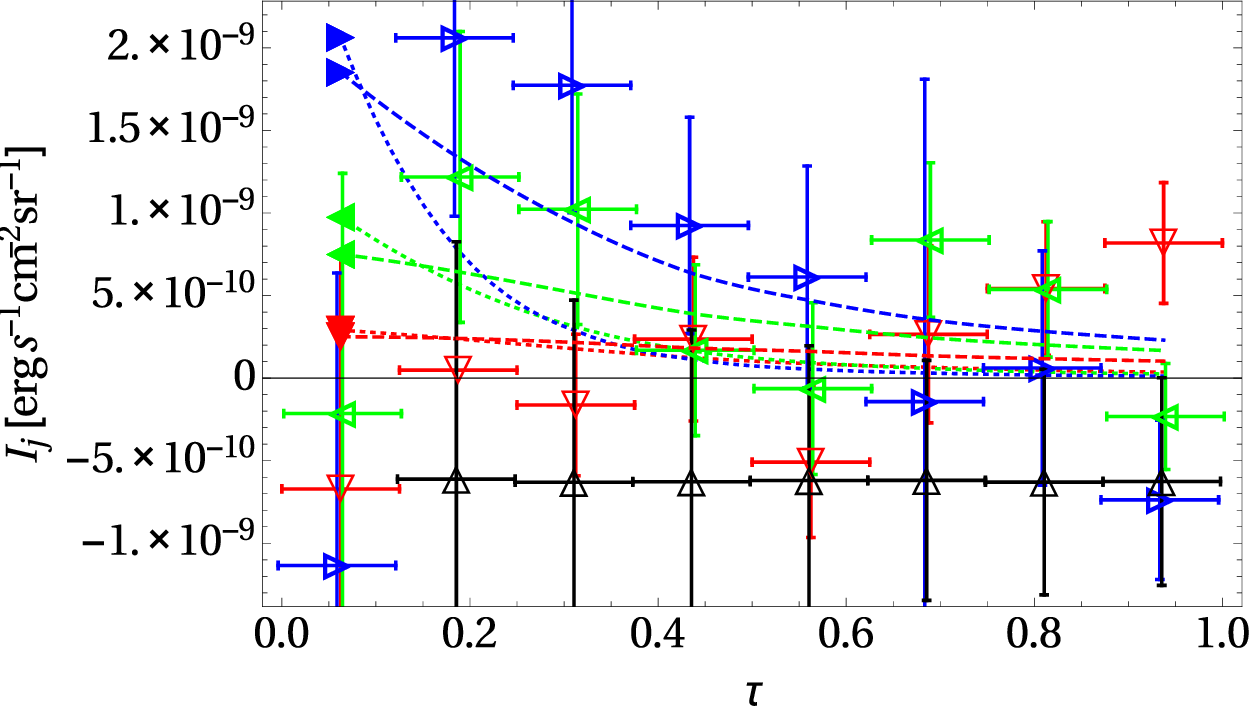}
    \vspace{-0.5cm}
   	\caption{
        Same as Fig.~\ref{fig:Flux} (top panel), but zooming into the central $\tau<1$ region, for the nominal sample ($\theta_{500}>0\dgrdot2$; left panel), the $\theta_{500}>0\dgrdot25$ sub-sample (31 non-compact clusters; middle), and the $\theta_{500}>0\dgrdot3$ sub-sample (18 extended clusters; right), along with best-fit point-source (dotted curves) and diffuse (dashed curves) nominal models (channels distinguished by colors and filled triangles in central bin).
        The high-energy channel is shown even where its photon statistics are insufficient for stacking, presenting as a fixed negative brightness (outside $\tau>0.5$ for the nominal sample, and for additional bins for the smaller, more extended samples).
    }
    \label{fig:Zoom}
\end{figure*}

We therefore focus on the $\tau<1$ region, and use a nominal $\Delta\tau=1/8$ bin size, to resolve the central signal; the resulting brightness profiles in each channel are shown in Fig.~\ref{fig:Zoom}.
In resemblance of the virial-shock signal of {\PapI}, the central signal too arises from the cumulative contributions of a large number of clusters, and is not dominated by a handful of centrally-bright cores.
The signal thus emerges independently in sub-samples obtained by introducing additional cuts, such as among clusters of low mass ($3.9\sigma$ for 41 clusters of $M_{500}<10^{14}M_\odot$) vs. high mass ($2.8\sigma$ for 34 clusters of $M_{500}>10^{14}M_\odot$), or in clusters of small ($2.3\sigma$ for 44 clusters of $\theta_{500}<0\dgrdot25$), intermediate ($4.1\sigma$ for 13 clusters of $0\dgrdot25\theta_{500}<0\dgrdot3$), or large ($1.8\sigma$ for 18 clusters of $\theta_{500}>0\dgrdot3$) angular scales.
In addition to the nominal sample (left panel), Fig.~\ref{fig:Zoom} also shows the brightness profiles stacked over the 31 clusters of $\theta_{500}>0\dgrdot25$ (henceforth `non-compact clusters'; middle panel) and the 18 clusters of $\theta_{500}>0\dgrdot3$ (henceforth `extended clusters'; right panel).
While the fewer clusters in the latter samples entail increasingly poor statistics, their larger angular extents better resolve the signal.

As the nominal analysis (left panel of Fig.~\ref{fig:Zoom}) shows, in the high energy bands 3 and 4 (blue right- and black up-triangles), and marginally in band 2 (green left-triangles), the central signal appears to be spatially extended.
However, due to the extended PSF and the small $\theta_{500}$ of many clusters, it is difficult to distinguish between extended emission vs. PSF-convolved emission from a central point source (dashed vs. dotted curves derived in \S\ref{sec:Modelling}) at a high confidence.
Nevertheless, when examining clusters of increasingly large $\theta_{500}$ (middle and right panels), the extended nature of the signal becomes evident.
Indeed, as shown in \S\ref{sec:Modelling}, the extent then becomes statistically significant, despite the smaller sample size, so a central point source can be ruled out.
Note that reducing the sample size renders the high-energy channel 4 unusable, as increasingly more bins have insufficient photon counts for statistical tests (these bins are pointed out in the figure as having a constant negative brightness).
Unlike {\PapI}, which reported the presence of unidentified central point sources, which were not yet detected by 3FGL, and quantified \eg their hard, $\mys\simeq 1.5$ spectrum, in the present analysis we find no evidence for central point sources undetected by 4FGL-DR4.

A backwards best-fit to the nominal $\tau<1$ results indicates a $dI(\epsilon=10\GeV)/d\ln\epsilon = 10^{-9.2\pm0.2}\erg\se^{-1}\cm^{-2}\sr^{-1}$ mean-cluster brightness, with a $\mys = 1.7\pm0.2$ photon spectral index.
Such emission, arising from the central $\tau<0\dgrdot5$ region, agrees with the $\sim10^{-9}\erg\se^{-1}\cm^{-2}\sr^{-1}$ brightness reported \citep{XiEtAl18_Coma, AdamEtAl21, BaghmanyanEtAl22} in Coma.
Such brightness lies a factor of $>4$ below previous upper limits \citep[\eg][]{AckermannEtAl10, GriffinEtAl14}, which were imposed on the central flux by stacking LAT data around clusters without $R_{500}$ rescaling.
Such a brightness level, the spatial extent of the signal, the comparable contributions made by a large number of clusters, and the spectral index, all seem consistent with the diffuse emission from CRI, and are less natural in alternative models.
In particular, emission from unresolved central point sources is disfavored, based on the above reasons, as well as the rigorous masking of all significant 4FGL point sources and the exclusion of all clusters with a nearby significant compact source.
However, robust estimates and model tests require forward modeling, which incorporates details such as the actual exposure and PSF distributions into well-defined models, addressed next.

\section{Point-source vs. hadronic models}
\label{sec:Models}

Consider first the simpler (and later disfavored) possibility that the central signal in Fig.~\ref{fig:Zoom} arises from a population of faint point sources in the centers of multiple clusters, each with a flux insufficient for its individual 4FGL detection.
For a source of specific luminosity $L_{\epsilon'}$ at an emitted photon energy $\epsilon'=(1+z)\epsilon$, the expected LAT specific brightness profile at some angular sky position $\bmt$ with respect to the center of the cluster is
\begin{equation}\label{eq:Psc}
  I_\epsilon(\bmt) = \frac{(1+z)L_{\epsilon'}}{4\pi d_L^2}P(\bmt,\epsilon) \, ,
\end{equation}
where $d_L(z)$ is the luminosity distance and $P(\bmt,\epsilon)\simeq P(\theta,\epsilon)$ is the LAT PSF toward the source position, defined normalized with unit integral over solid angle $\Omega$,
\begin{equation}\label{eq:PSF}
  \int P(\bmt,\epsilon)\, d\boldsymbol{\Omega} = 1 \, .
\end{equation}
We consider different simplified distributions of such point sources, including the nominal case where all sources have the same $F_\epsilon'=(1+z)L_{\epsilon'}/4\pi d_L^2$ specific flux, motivated by some 4FGL sensitivity limit.
Also considered, in particular, is the limit where all sources have the same intrinsic luminosity $L_{\epsilon'}$.
For simplicity, we approximate the radiated energy spectrum $L_{\epsilon}\propto \epsilon^{-q}$ as a power law of index $q$ for observed photons in the LAT band; in general, $q\neq \mys$ because the PSF is broad and varies with angle, energy, and sky position.

Next, consider the hadronic signature of a CRI population of differential number-density distribution $dN(\bmr,E)/dE$, inelastically colliding with ambient ICM nuclei of number density $n(\bmr)$, where $\bmr$ and $E$ are the CRI position and energy.
The resulting (hadronic model, henceforth) LAT signal is given by
\begin{equation}\label{eq:HModel}
  \!\!\!I_\epsilon(\bmt) = \! \frac{c \epsilon'}{4\pi(1+z)^3} \!\! \int\!\! d\boldsymbol{\Omega} \, P(\bmp) \!\! \int\!\! dl \, n(\bmr) \!\! \int \!\! dE \frac{dN(\bmr,E)}{dE}\,  \frac{d\sigma_\gamma(E,\epsilon')}{d\epsilon'}\coma \!\!
\end{equation}
where
$\bmp$ is the angular separation between the $\bmt$ and $\boldsymbol{\Omega}$ sky directions,
$l$ is the line-of-sight coordinate,
and
$c$ is the speed of light.
The differential inclusive cross-section
$d\sigma_\gamma(E,\epsilon')/d\epsilon'$ for $\epsilon'$ photon production is computed using the formulation of \citet{KamaeEtAl06},
with corrected parameters and cutoffs (T. Kamae \& H. Lee, private communications 2010).

We approximate the CRI energy spectrum as a power-law $dN/dE \propto E^{-p}$ of index $p$.
A photon produced from an inelastic $p$-$p$ collision carries on (energy-weighted) average $\sim10\%$ of the parent CRI energy $E$, so for convenience we normalize the CRI distribution at $E=100\GeV$, roughly corresponding to the central $\epsilon=10\GeV$ photon energy in the analyzed LAT band.
The CRI spatial distribution is poorly constrained a-priori, as it depends on the CRI accelerators and on their subsequent evolution by advection and diffusion through the ICM, so we parameterize $dN/dE \propto n(\bmr)^\sigma$ as a simple function of the ambient nucleon density.
The power-law index $\sigma$ is often taken as unity, assuming that CRI follow the gas, but in present notations it would be $2/3$ if CRI are only advected from the virial shock \citep{KushnirWaxman09}, or zero if, as indicated by radio observations in the hadronic model \citep{Keshet10, Keshet24} and the LAT signal in Coma \citep{Keshet24, KushnirEtAl24}, strong diffusion or mixing homogenizes the CRI.

In some model variants below, the central gas and CRI densities diverge, so a finite reference radius is useful for quantifying the model.
Different studies find the scaled core radii of low-redshift clusters to be on average $\tau_c\lesssim 0.1$ \citep[\eg][]{HudsonEtAl10}, $\sim 0.1$ \citep{Mohr99, EckertEtAl12}, or $\sim0.2$ \citep{KaferEtAl19}, and some studies consider the range $0.1$--$0.3$ \citep{ArnaudEtAl10}.
More importantly, among our sample clusters with available core measurements (see \ref{sec:Sample}), the mean value is $\tau_c\simeq 0.10$ and the median is $\tau_c=0.08$; these results hold both for the present nominal 75-cluster sample and for the larger, 112-cluster sample of {\PapI}.
Therefore, and as the results are found below to be sensitive to $dN/dE$ around this length scale, we choose for convenience $\tau=0.1$ as a reference radius.
The above arguments. along with the simplifying assumption of spherical symmetry, then motivate the differential CRI energy-density parametrization
\begin{equation}\label{eq:CRI}
  \frac{du(\bmr,E)}{d\ln E} \equiv E^2 \frac{dN}{dE} = A \, E_{100}^{2-p} \left[ \frac{n\left(r\right)}{n\left(0.1R_{500}\right)}\right]^\sigma  \coma
\end{equation}
where we defined $E_{100}\equiv E/100\GeV$ and the normalization $A\equiv du(\tau=0.1,E=100\GeV)/d\ln E$.

In order to fully define the hadronic model, we now need only specify the functional form of the spatial gas distribution $n(r)$; its normalization is then dictated for each cluster by the MCXC-tabulated $M_{500}$ and $R_{500}$ values.
Studies based on \emph{ROSAT}/PSPC \citep{VikhlininEtAl99, EckertEtAl12} suggest approximate self-similarity for $\tau\gtrsim 0.3$, with the inner profile approximately following a $\beta$-model with $\beta$ in the range $0.65\mbox{--}0.85$, but steepening beyond $R_{500}$ to somewhat larger $\beta$ values ranging from $\beta\simeq 0.8$ in unrelaxed clusters to $\beta\simeq 1$ in cool-core clusters. In such cool-core clusters, a cusp should be taken into account at small radii \citep[\eg][]{PrattArnaud02, SandersonEtAl09, HudsonEtAl10}.

As our analysis is based on the MCXC, we nominally adopt the same default parametrization, itself based on the \citet{PrattArnaud02} AB model
\begin{equation}\label{eq:nMCXC}
  n\propto \left[1+\left({\tau}/{\tau_c}\right)^2\right]^{-\frac{3\beta}{2}+\frac{\alpha}{2}} \tau^{-\alpha} \coma
\end{equation}
with $\beta=0.768$ and $\alpha=0.525$ \citep{PiffarettiEtAl11}.
We nominally adopt a scaled $\tau_c=0.1$ core radius, based on the aforementioned models available for specific clusters in our sample, but consider also values as large as $\tau_c=0.3$.
We also consider alternative gas distributions including $\beta$-models without a cusp ($\alpha=0$), AB or $\beta$ models with steepening beyond $\tau=1$, and AB models with an NFW-like ($\alpha=1$) cusp.
In general, the results do not strongly depend on the values of $\alpha$ and $\beta$, but are somewhat sensitive to $\tau_c$.

\section{Best fit: flat CRI distribution}
\label{sec:Modelling}

We examine each of the above model variants by injecting the $I_\epsilon(\bmt)$ distribution it implies for each cluster, after convolution with the coincident PSF, into its surrounding non-masked HEALPix pixels, and then bin and stack the results over the sample clusters using the same pipeline applied to the real data.
Next, linear regression yields best-fit model-parameter values and uncertainties (estimate standard errors), and the goodness of fit is evaluated from the chi-squared value $\chi^2$ per number $\nu$ of degrees of freedom.
We also examine the TS-test, where $\mbox{TS} = \chi^2_- - \chi^2_+$ compares $\chi^2$ values for models of $\mathsf{n}$ parameters, obtained before ($-$ subscripts) and after ($+$ subscript) adding the modelled central component to the best-fitted field; TS then approximately follows a chi-squared distribution $\chi_\mathsf{n}^2$ of order $\mathsf{n}\equiv \mathsf{n}_+-\mathsf{n}_-$ \citep{Wilks1938}.

The resulting best-fit profiles are presented in Fig.~\ref{fig:Zoom} for the point source (dotted curves) and hadronic (dashed) nominal models.
The nominal 75-cluster sample has sufficient statistics for fitting all four energy channels, providing a robust $1.9\pm0.4$ (multivariate) spectral-index fit that remains unchanged across all hadronic (for $p$, with two additional parameters $A$ and $\sigma$) and point-source (for $q$, with one additional parameter, $F_\epsilon$ or $L_\epsilon$) models variants.
However, due to the compact, $\theta_{500}<0\dgrdot25$ clusters included in this sample, the hadronic model (projected $\chi^2/\nu=0.90$ and TS $4.1\sigma$ for $\mathsf{n}=2$) provides only a marginally better fit than the point-source model ($\chi^2/\nu=0.94$; TS $3.9\sigma$ for $\mathsf{n}=2$).
Here, in order to compare models with the same number of free parameters, we first projected the hadronic model onto its best-fit $\sigma$ (or $p$ or $A$) value; otherwise, the $\mathsf{n}=3$ metrics ($\chi^2/\nu=0.93$; TS $3.7\sigma$) of the nominal-sample fit are not necessarily better than those of the $\mathsf{n}=2$ point-source model.

The point-source and hadronic models are better distinguished using a higher effective resolution, achieved by excluding compact clusters.
The high-energy channel 4 is still useful for the non-compact sub-sample (31 clusters of $\theta_{500}>0\dgrdot25$) at low, $\Delta\tau=1/4$ resolution, indicating that $p=1.99\pm0.42$; in this case the extent parameter has a small effect, so we may project onto its best fit to obtain
\begin{equation}\label{eq:Bestp}
  p=1.99\pm0.32 \, .
\end{equation}
For higher resolutions or smaller sub-samples, the underlying $\Myf\gg1$ assumption is in general no longer satisfied.
Therefore, we avoid channel 4 when using the non-nominal sub-samples to study the spatial distribution.

For the 31-cluster sub-sample, the hadronic model ($\sigma$-projected $\chi^2/\nu=0.95$ and TS $3.6\sigma$ for $\mathsf{n}=2$, or $\chi^2/\nu=1.00$ and TS $3.3\sigma$ for $\mathsf{n}=3$) then provides a noticeably better fit than the corresponding point-source model ($\chi^2/\nu=1.28$; TS $2.5\sigma$).
As expected, this trend becomes stronger when considering the smaller, extended sub-sample (18 clusters of $\theta_{500}>0\dgrdot3$), where the hadronic model (projected $\chi^2/\nu=1.06$ and TS $2.7\sigma$ for $\mathsf{n}=2$, or $\chi^2/\nu=1.11$ and TS $2.3\sigma$ for $\mathsf{n}=3$) is still acceptable, given the reduced number of clusters, whereas the point-source model ($\chi^2/\nu=1.33$; TS $1.4\sigma$) is rejected.
Joint fits combining both hadronic and point-source emission favor a negligible or exceedingly soft point-source component, in all samples.

\begin{figure}[t!]
    \vspace{-.3cm}
    \begin{center}
        \begin{tikzpicture}
            \draw (0, 0) node[inner sep=0]
            {
                \includegraphics[width=0.38\textwidth, trim={0 0.05cm 0 0.05cm}, clip]{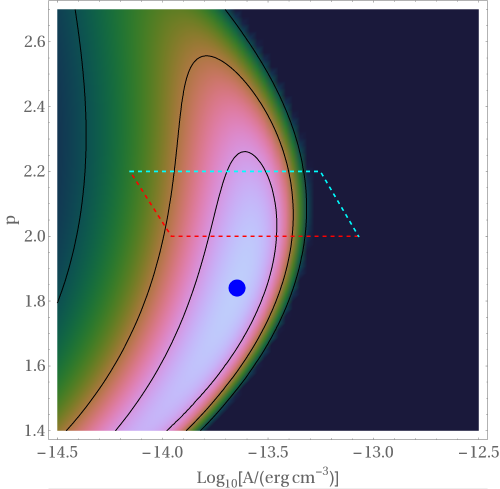}
            };
            \draw (1.9, -2.2) node[text=white] {\scriptsize $\sigma_{j=1..3}=0.10\pm0.48$};
       \end{tikzpicture}
    \end{center}
    \vspace{-0.5cm}
    \begin{center}
        \begin{tikzpicture}
            \draw (0, 0) node[inner sep=0]
            {
                \hspace{-0.4cm}
                \includegraphics[width=0.38\textwidth, trim={0 0.05cm 0 0.03cm}, clip]{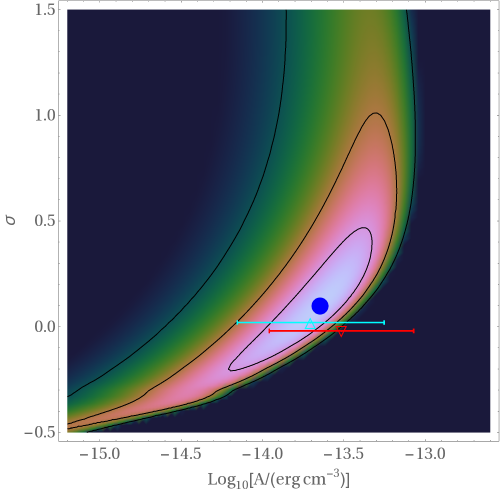}
            };
            \draw (2.1, -2.3) node[text=white] {\scriptsize $p_{j=1..3}=1.84\pm0.64$};
       \end{tikzpicture}
    \end{center}
    \vspace{-0.5cm}
    \begin{center}
        \begin{tikzpicture}
            \draw (0, 0) node[inner sep=0]
            {
                \includegraphics[width=0.375\textwidth, trim={0 0.05cm 0 0.05cm}, clip]{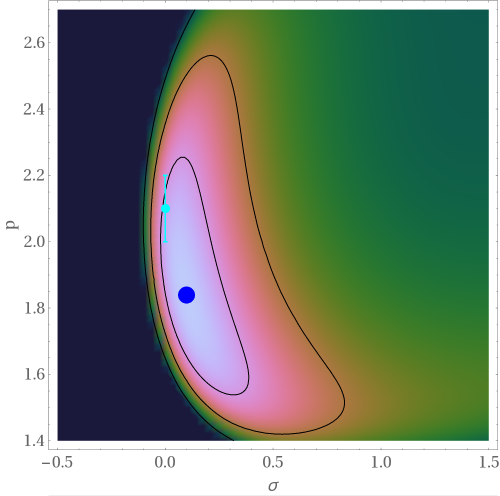}
            };
            \draw (2.1, -2.3) node[text=white] {\scriptsize $A_{j=1..3}=10^{-13.6\pm0.8}$};
       \end{tikzpicture}
        \vspace{-0.6cm}
    \end{center}
   	\caption{
        Orthogonal projections (see labels) of best-fit (blue disk with $1\sigma$, $2\sigma$, and $3\sigma$ multi-variate confidence-level contours) $\mathsf{n}=3$ CRI parameters in the nominal analysis (without channel $j=4$) of the non-compact ($\theta_{500}>0\dgrdot25$ cluster) sub-sample.
        Predictions based on a joint compilation of radio MHs, GHs, and relics \citep{Keshet10} are also shown (parallelogram or error bars for $p=2.0, 2.2$ as down, up triangles with $\sigma=0$ slightly offset vertically for visibility).
    }
    \label{fig:ModelContours}
    \vspace{-1.cm}
\end{figure}
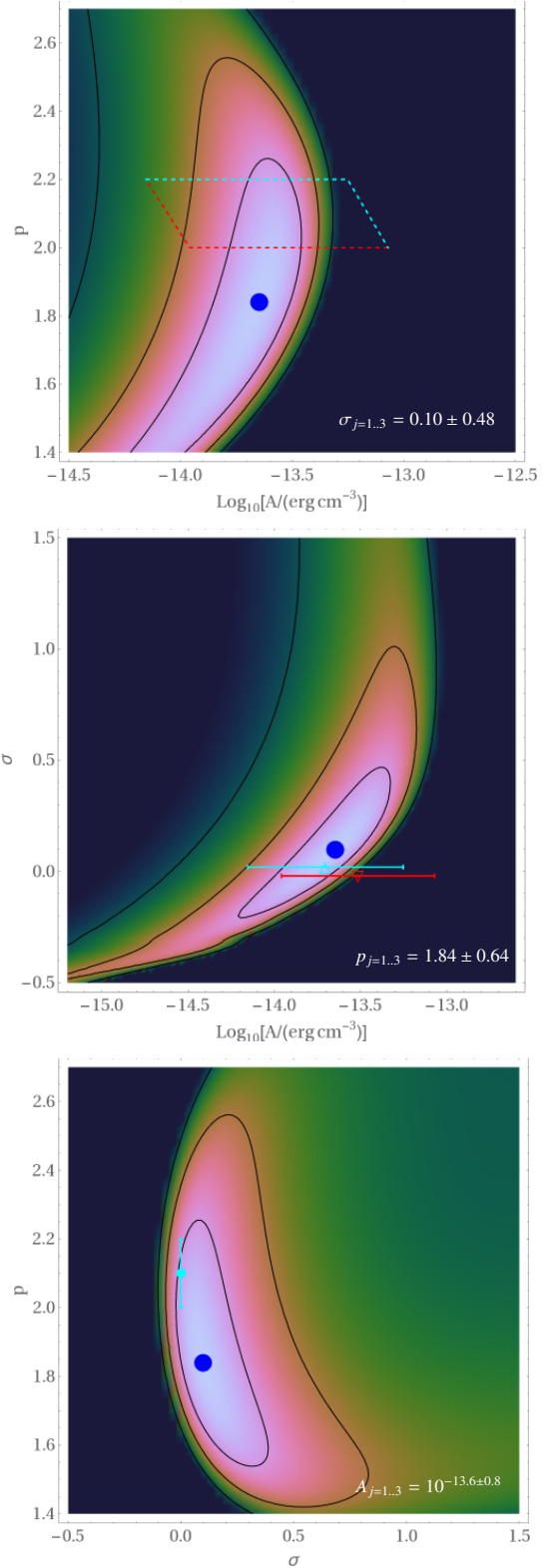

The same difficulty in resolving the emission from compact clusters also precludes a determination of the CRI spatial extent parameter $\sigma$ in the nominal sample, so we focus on more extended clusters.
In particular, we use the sub-sample of 31 non-compact clusters, taking advantage of its significant ($4.2\sigma$) central excess, to study the three-channel hadronic model.
Figure \ref{fig:ModelContours} presents the resulting confidence-level contour plots for the $\mathsf{n}=3$ hadronic model, projected at the best fit (see labels) $\sigma$ (top panel), $p$ (middle), or $A$ (bottom) values.
The spatial index $\sigma=0.10\pm0.48$ indicates an extended CRI distribution, consistent with the success of the hadronic model with respect to the point-source model.
As $\sigma$ depends weakly on the spectral index, we may project onto the best-fit $p$ to obtain
\begin{equation}\label{eq:BestSigma}
  \sigma=0.10\pm0.36\, .
\end{equation}
The smaller sub-sample of 18 extended clusters confirms these results, but with larger uncertainties, giving
$\sigma=0.06\pm0.67$ (or $\sigma=0.06\pm0.51$ for projected $p$).

The inferred hadronic model parameters do not strongly depend on analysis parameters
(see \ref{sec:Sensitivity}) or on variations in model parameters (see \ref{sec:Variants}).
Varying the $\alpha$ or $\beta$ parameters of the gas distribution does not significantly change $p$ and only mildly changes $\sigma$; the inferred CRI distribution remains highly extended even if $n(r)$ has no central cusp ($\sigma=0.14\pm0.50$ for $\alpha=0$) or is shallow ($\sigma=0.26\pm0.54$ for $\beta=2/3$).
In the opposite extremes of an NFW-like ($\alpha=1$) cusp or a steep ($\beta=1$) profile, the inferred CRI are more homogeneous ($\sigma\simeq 0$).
Incorporating a steepening in the gas profile beyond $\tau=1$ has a negligible effect on the results.
The inferred CRI distribution is somewhat more sensitive to changes in the core radius, giving $\sigma=0.49\pm0.82$ if one adopts $\tau_c=0.3$, but such large cores are not consistent with our sample;
note that measured core radii in the non-compact or extended sub-samples have even smaller cores than in the nominal sample.

The best-fit $A=10^{-13.6\pm0.8}\erg\cm^{-3}$ ($\mathsf{n}=3$ non-projected) normalization is obtained robustly, consistent across all model variants.
Inspection of Fig.~\ref{fig:ModelContours} indicates that the large uncertainty in this multivariate fit arises from phase-space regions that are non-physical in a CRI model, such as spatial $\sigma<0$ indices or energy spectra much harder than $p=2$.
Again, $A$ is not sensitive to $p$, so a conservative estimate is obtained by projecting onto the best-fit (or any reasonable) $p$, giving
\begin{equation}\label{eq:BestA}
  A=10^{-13.6\pm0.5}\erg\cm^{-3} \, .
\end{equation}
Tighter estimates are obtained by marginalizing over the allowed phase space or projecting onto the best-fit $\sigma$ ($A=10^{-13.6\pm0.3}\erg\cm^{-3}$); in the extreme, projecting onto both $p$ and $\sigma$ yields $A=10^{-13.6\pm0.1}\erg\cm^{-3}$.

Predictions for the hadronic model \citep{Keshet10} based on diffuse intracluster radio emission in its various forms are presented in Fig.~\ref{fig:ModelContours} (and in subsequent confidence-level plots, as a dashed parallelogram or error bars).
Focusing on the non-compact sub-sample, which shows a significant central excess and has sufficient angular resolution to resolve the CRI distribution, the best-fit $p=2.0\pm0.4$ spectral index is consistent with the $2.0<p<2.2$ prediction, and so is the extended, nearly homogeneous spatial index $\sigma=0.1\pm0.3$.
More importantly, the CRI energy density $A$ in Eq.~\eqref{eq:BestA} agrees well with its prediction in Eq.~\eqref{eq:CRI_prediction}, indicating that the same CRI responsible for the \gama-ray excess are also those driving the diffuse ICM radio emission in its various forms.

\section{Summary and discussion}
\label{sec:Discussion}

This Paper II of the series, dedicated to stacking \emph{Fermi}-LAT data scaled around MCXC clusters, updates {\PapI} with the most recent, 4FGL-DR4 LAT source catalog, leading to a high-quality sample of 75 high-latitude, $\theta_{500}>0\dgrdot2$ clusters (and a sub-sample of 31 non-compact, $\theta_{500}>0\dgrdot25$ clusters), presented as disks (dark, non-cyan disks) in Fig.~\ref{fig:Sample75}, and focuses on the significant, $4.7\sigma$ ($4.2\sigma$) stacked excess in their central, $\tau<0.5$ region; see Figs.~\ref{fig:VSummary} and \ref{fig:Flux}.
This robust (see \S\ref{sec:CentralExcess} and \ref{sec:Sensitivity}) $1$--$100$ GeV excess, consistent with recent measurements in Coma, is a factor of $>4$ lower than previous upper limits on the central flux, imposed by stacking analyses \citep[\eg][]{AckermannEtAl10, GriffinEtAl14} that did not scale clusters to their natural, \eg $R_{500}$ scales prior to stacking.
The central excess is spatially extended and inconsistent with central AGN (Fig.~\ref{fig:Zoom}); it is best fit (Fig.~\ref{fig:ModelContours} and its variants in \ref{sec:Variants}) as hadronic emission \eqref{eq:HModel} from a CRI distribution \eqref{eq:CRI} of flat energy spectral index \eqref{eq:Bestp}, a nearly homogeneous spatial index \eqref{eq:BestSigma}, and the implied $\sim$constant differential energy density \eqref{eq:BestA}. 
Including poorly resolved clusters in the sample tends to raise $\sigma$ and $p$.

An alternative model of central point sources is disfavored, and a distribution of point sources mimicking hadronic emission is unlikely, for several reasons:
(i) Sample clusters are carefully selected to avoid any cataloged point source in their center;
(ii) Significant ($>5\sigma$) sources are masked within the $95\%$ LAT containment;
(iii) Clusters are furthermore selected to avoid any well-localized ($a_{95}<0\dgrdot08$) source within a large, $1\dgrdot8$ radius (giving $95\%$ containment at the lowest energy);
(iv) The central excess is spatially extended and approximately follows the gas density;
(v) The signal is spectrally flatter than typical AGN sources;
and
(vi) The signal emerges as the cumulative sum of small contributions from many clusters, so cannot be attributed to few unidentified sources that escaped detection.
For additional evidence, see Paper III.

The properties of the inferred CRI distribution are in good agreement with the prediction \eqref{eq:CRI_prediction}, which is based on jointly attributing the diffuse ICM radio emission in its various forms \citep[minihalos, giant halos, and relics, their transitional forms, and the recently discovered mega-halos;][]{Keshet10, Keshet24} to the same $p\simeq 2$, approximately homogeneous CRI distribution.
This agreement supports the hadronic interpretation of both the present central \gama-ray excess, and the diffuse ICM radio emission.
Indeed, the inferred CRI energy density \eqref{eq:BestA}, sufficient to account for the radio emission, leaves little room for leptonic models that invoke primary electron acceleration in weak shocks or turbulence; for discussion and additional evidence for the joint hadronic model, see \citet{Keshet24}.
Our results also agree with the recent \gama-ray signals detected in Coma, provided that the CRI spectral index is not assumed soft \citep{Keshet24, KushnirEtAl24}.

Attributing the detected \gama-ray central excess to hadronic emission strengthens previous upper limits on any residual $1$--$100$ GeV emission, which might arise from alternative processes such as dark matter annihilation.
We thus impose an upper limit $10^{-9}\erg\se^{-1}\cm^{-2}\sr^{-1}$ on the brightness of such a component.
Attempts to improve the model by combining hadronic emission with a central point-source component attribute a vanishing amplitude or an exceedingly soft spectrum to the source, so its upper limit is on the order of the hadronic central-flux uncertainty.
For the $\sim0\dgrdot1$ PSF $68\%$ containment in the high-energy channels, we thus obtain a central $dF/d\ln\epsilon < 10^{-14}\erg\se^{-1}\cm^{-2}$ upper limit.

Finally, note that \gama-ray emission from charged pions was neglected here, as well as in most previous estimates of the hadronic \gama-ray signal.
Such emission is always weaker than the $\pi^0\to\gamma\gamma$ signal, by at least a factor of $2$ for a flat CRI spectrum \citep{KamaeEtAl06}, and becomes negligible in highly magnetized regions such as cluster cores.
This neglected component would not significantly modify any of the above conclusions, even if the ambient magnetic field in the center were weak.

\paragraph*{Acknowledgements} \begin{small}
I am grateful to I. Reiss, K.~C. Hou, I. Gurwich, Y. Lyubarsky, and the late G. Ilani, for helpful discussions.
This research was supported by the Israel Science Foundation (ISF grant No. 2126/22).
\end{small}

\bibliographystyle{elsarticle-harv-author-truncate}
\bibliography{Virial}

\appendix

\section{Sensitivity tests}
\label{sec:Sensitivity}

Our results are not sensitive to the precise choice of analysis parameters and selection cuts, as demonstrated above and shown in detail in {\PapI}.
Most of the convergence and sensitivity tests of {\PapI} are valid for the present analysis as well, so we focus on the aspects specific to the latter.

The present analysis introduces two new parameters, controlling the $\sim5\sigma$ significance and $a_{95}\sim0\dgrdot08$ localization selection criteria  of 4FGL-DR4 sources used for source masking and cluster exclusion.
However, the results are not sensitive to the values of these parameters, as long as the two sources located in the centers of high-latitude clusters are masked, or the two host clusters are excluded from the sample.

\begin{figure}[h]
    \centering
    \includegraphics[width=0.45\textwidth]{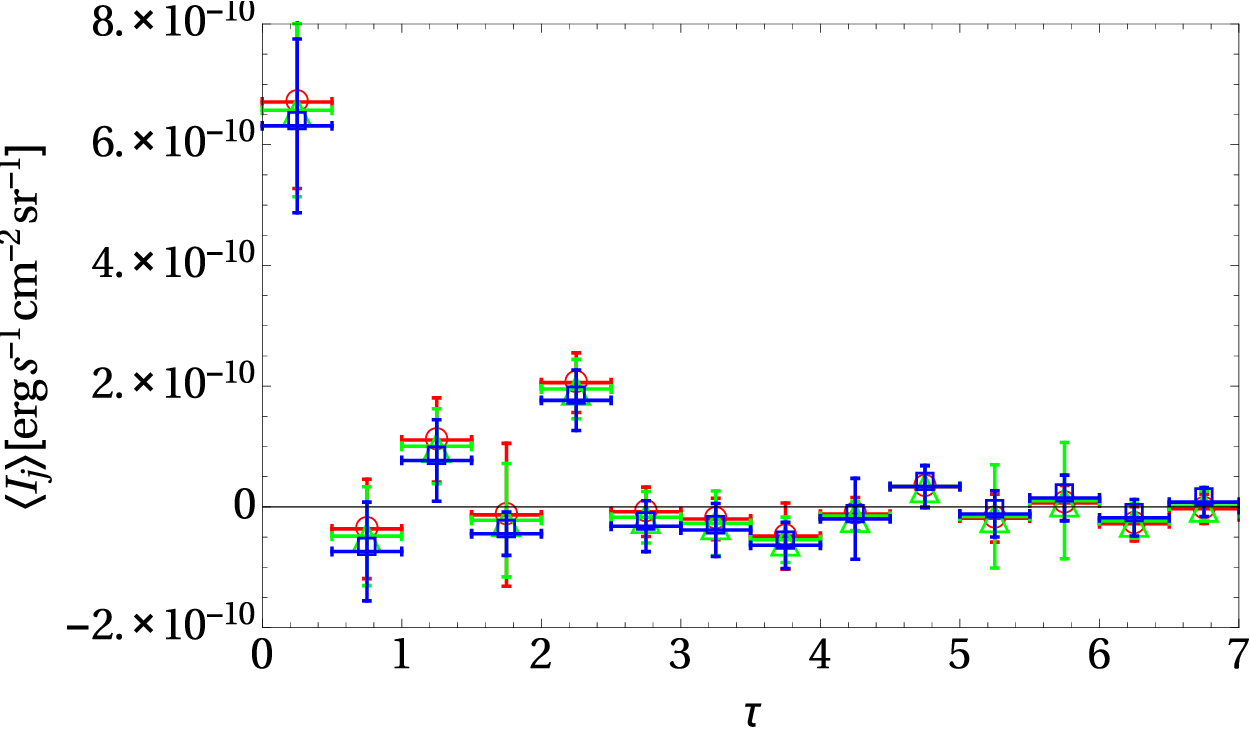}
    \vspace{-0.3cm}
   	\caption{
        Same as Fig.~\ref{fig:Flux} (its bottom panel for epoch I with $\Delta\tau=1/2$ resolution), but for foreground orders $N_f=0$ (red circles), 2 (green triangles), and 4 (blue squares).
    }
    \label{fig:SensFOrder}
\end{figure}

The order $N_f$ of the foreground fit could in principle change the field level interpolated to the center of the cluster, with some impact on the our results.
However, as Fig.~\ref{fig:SensFOrder} demonstrates, the effect is quite weak.
For instance, the central signal is nominally $4.7\sigma$ ($4.4\sigma$) for $N_f=0$ ($N_f=4$).
We confirm that the best-fit parameters are not sensitive to the order $N_f$ or the $\Delta\tau\lesssim 1/4$ resolution.

\section{LAT PSF in epoch I vs. II }
\label{sec:EmpiricalPSF}

The \texttt{gtpsf} function in \texttt{fermi tools} is used to model the PSF, including its dependence on photon energy, sky position, and other observational parameters, but we also empirically test the LAT PSF and sensitivity for any possible temporal evolution.
For this purpose, we stack the brightness distribution around the brightest, most compact 3FGL sources found in low foreground regions.
In particular, we select the 15 significant, $>12\sigma$ sources of high, $|b|>40\dgr$ latitudes, $a_{95}<0\dgrdot01$ semi-major axes, and   $>10^{-8} \se^{-1}\cm^{-2}$ flux, listed in Table \ref{tab:psc}.

\begin{table}[h!]
{\small %
    	\caption{
            \label{tab:psc}
    		Point sources used for PSF/sensitivity evolution test.\\
            \textbf{Columns:} (1) 3FGL source name; (2) Latitude (deg); (3) Longitude (deg);
            (4) $1$--$100$ GeV flux in s$^{-1}\cm^{-2}$ units.
        }
        \vspace{0.3cm}
{\small
        \begin{tabular}{lccc}
    		Source Name &  $b$ & $l$ & $F(1\text{--}100\GeV)$ \\
    		(1) & (2) & (3) & (4) \\
    		\hline
            3FGL J1512.8-0906  & $40.13$ & $351.29$ & $4.11\times 10^{-8}$\\
            3FGL J1104.4+3812  & $65.03$ & $179.83$ & $3.03\times 10^{-8}$\\
            3FGL J1224.9+2122  & $81.66$ & $255.06$ & $2.55\times 10^{-8}$\\
            3FGL J1504.4+1029  & $54.58$ & $11.38$ & $2.40\times 10^{-8}$\\
            3FGL J2158.8-3013  & $-52.25$ & $17.73$ & $2.17\times 10^{-8}$\\
            3FGL J1256.1-0547  & $57.06$ & $305.10$ & $2.06\times 10^{-8}$\\
            3FGL J0428.6-3756  & $-43.62$ & $240.70$ & $2.01\times 10^{-8}$\\
            3FGL J1522.1+3144  & $57.02$ & $50.16$ & $1.79\times 10^{-8}$\\
            3FGL J1231.2-1411  & $48.38$ & $295.54$ & $1.78\times 10^{-8}$\\
            3FGL J2329.3-4955  & $-62.31$ & $332.01$ & $1.41\times 10^{-8}$\\
            3FGL J1427.0+2347  & $68.20$ & $29.49$ & $1.36\times 10^{-8}$\\
            3FGL J1555.7+1111  & $43.96$ & $21.92$ & $1.32\times 10^{-8}$\\
            3FGL J1635.2+3809  & $42.34$ & $61.11$ & $1.14\times 10^{-8}$\\
            3FGL J2345.2-1554  & $-70.99$ & $65.70$ & $1.14\times 10^{-8}$\\
            3FGL J0957.6+5523  & $47.93$ & $158.59$ & $1.04\times 10^{-8}$\\
        \end{tabular}
}
}
\end{table}

\begin{figure}[t!]
    \centering
    \includegraphics[width=0.45\textwidth]{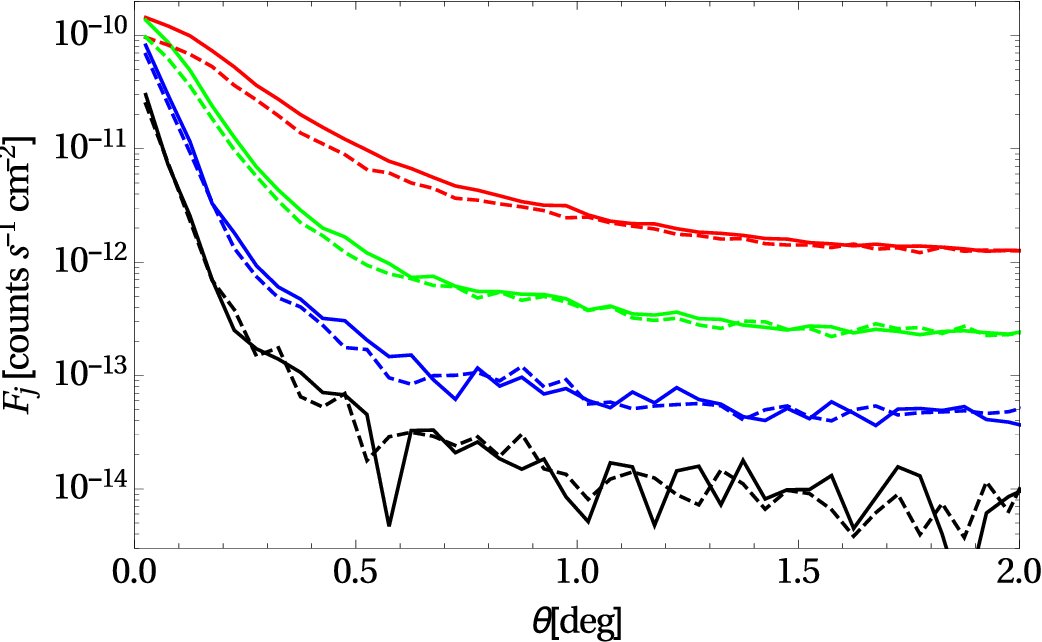}
    \vspace{-0.3cm}
    \caption{
    Photon flux radially binned and stacked around 15 high-latitude, bright and significant 3FGL sources, in early (LAT weeks 9-422; solid curves) vs. late (weeks 423-859; dashed) LAT epochs, in energy bands 1--4 (top to bottom). As long as sufficient sources of comparable flux are used, the result does not strongly depend on the specific choice of sources.
    }
    \label{fig:PSFMonitor}
\end{figure}

Although the overall results indicate good stability over the first $\sim8$ years of the mission, in agreement with the 10-year LAT performance analysis \citep{AjelloEtAl21}, we find evidence for some decline in localization after $\sim2016$, as illustrated in Fig.~\ref{fig:PSFMonitor}. As our results heavily rely on the good localization of high-energy photons, we thus chose to focus on the same 2008--2016 epoch of stable PSF studied in {\PapI}, albeit with newly downloaded data, tools, and recommended cuts.

\section{Model variants}
\label{sec:Variants}

\begin{figure*}[!b]
    \begin{center}
        \begin{tikzpicture}
            \draw (0, 0) node[inner sep=0]
            {
                \includegraphics[height=0.33\textwidth]{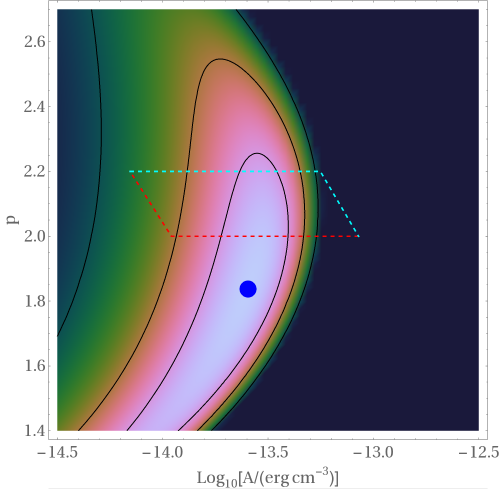}
                \includegraphics[height=0.33\textwidth]{Figures/ContR01th025SigBst.png}
                \includegraphics[height=0.33\textwidth]{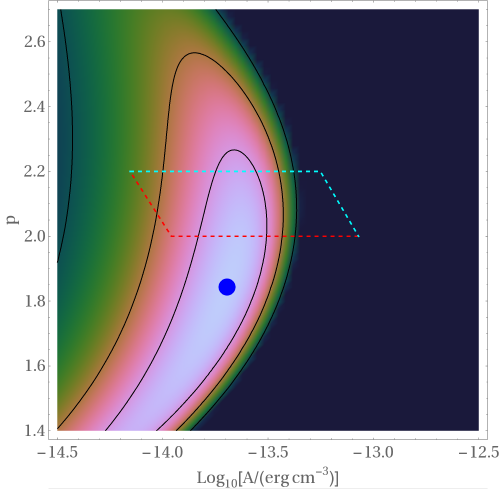}
            };
            \draw (-4.7, -2.0) node[text=white] {\scriptsize $\sigma_{j=1..3}=0.14\pm0.51$};
            \draw (1.7, -2.0) node[text=white] {\scriptsize $\sigma_{j=1..3}=0.10\pm0.48$};
            \draw (7.9, -2.0) node[text=white] {\scriptsize $\sigma_{j=1..3}=0.06\pm0.44$};
       \end{tikzpicture}
    \end{center}
    \begin{center}
        \begin{tikzpicture}
            \draw (0, 0) node[inner sep=0]
            {
                \includegraphics[height=0.33\textwidth]{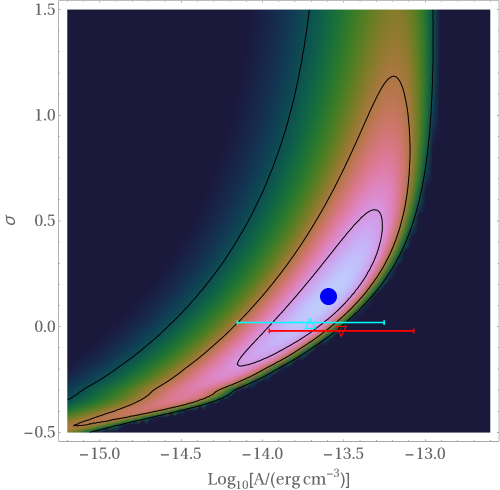}
                \includegraphics[height=0.33\textwidth]{Figures/ContR01th025pBst.png}
                \includegraphics[height=0.33\textwidth]{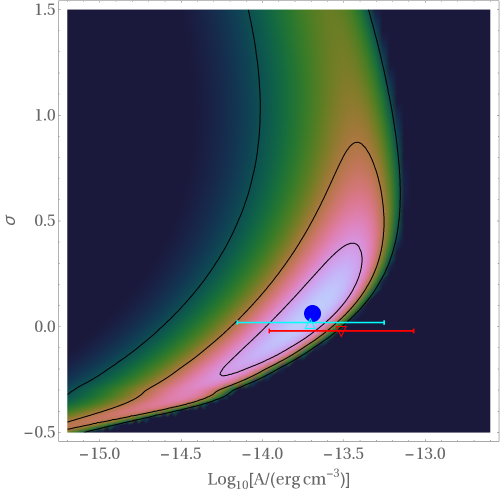}
            };
            \draw (-4.7, -2.0) node[text=white] {\scriptsize $p_{j=1..3}=1.84\pm0.64$};
            \draw (1.7, -2.0) node[text=white] {\scriptsize $p_{j=1..3}=1.84\pm0.64$};
            \draw (7.9, -2.0) node[text=white] {\scriptsize $p_{j=1..3}=1.84\pm0.64$};
       \end{tikzpicture}
    \end{center}
    \begin{center}
        \begin{tikzpicture}
            \draw (0, 0) node[inner sep=0]
            {
                \hspace{0.1cm}
                \includegraphics[height=0.33\textwidth]{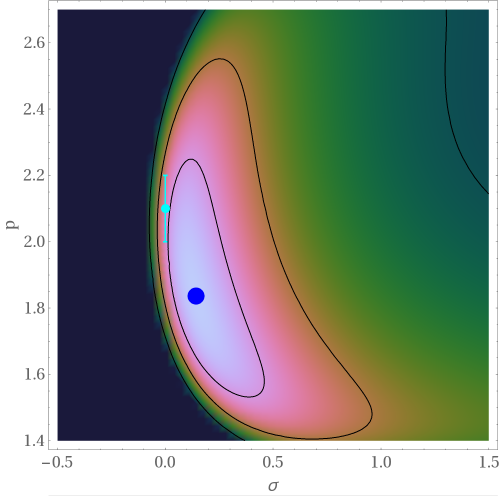}
                \includegraphics[height=0.33\textwidth]{Figures/ContR01th025ABst.png}
                \includegraphics[height=0.33\textwidth]{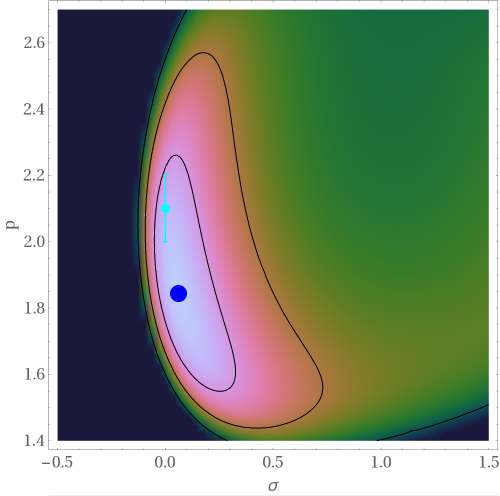}
            };
            \draw (-4.5, -2.0) node[text=white] {\scriptsize $A_{j=1..3}=10^{-13.6\pm0.8}$};
            \draw (1.7, -2.0) node[text=white] {\scriptsize $A_{j=1..3}=10^{-13.6\pm0.8}$};
            \draw (7.9, -2.0) node[text=white] {\scriptsize $A_{j=1..3}=10^{-13.7\pm0.8}$};
       \end{tikzpicture}
        \vspace{-0.6cm}
    \end{center}
   	\caption{
        Same as Fig.~\ref{fig:ModelContours} but varying the cusp profile: $\alpha=0$ (left column), $0.525$ (nominal; middle column), and $1$ (right column).
    }
    \label{fig:ModelContoursAlpha}
\end{figure*}

\begin{figure*}
    \begin{center}
        \begin{tikzpicture}
            \draw (0, 0) node[inner sep=0]
            {
                \includegraphics[height=0.33\textwidth]{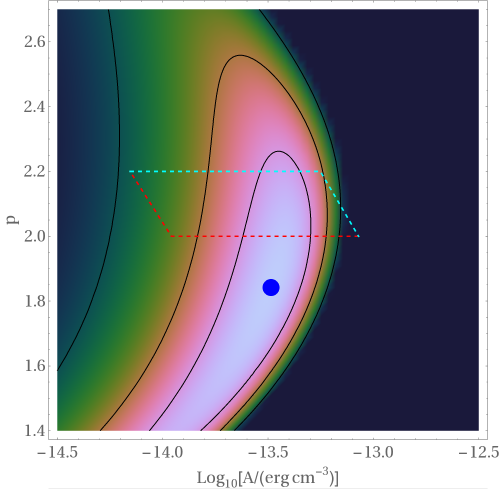}
                \includegraphics[height=0.33\textwidth]{Figures/ContR01th025SigBst.png}
                \includegraphics[height=0.33\textwidth]{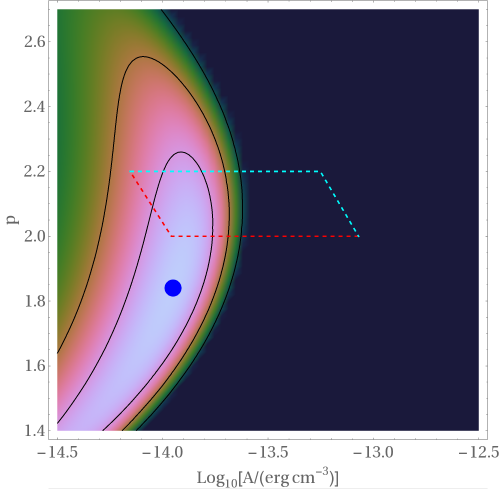}
            };
            \draw (-4.7, -2.0) node[text=white] {\scriptsize $\sigma_{j=1..3}=0.26\pm0.54$};
            \draw (1.5, -2.0) node[text=white] {\scriptsize $\sigma_{j=1..3}=0.10\pm0.48$};
            \draw (7.7, -2.0) node[text=white] {\scriptsize $\sigma_{j=1..3}=-0.15\pm0.36$};
       \end{tikzpicture}
    \end{center}
    \begin{center}
        \begin{tikzpicture}
            \draw (0, 0) node[inner sep=0]
            {
                \includegraphics[height=0.33\textwidth]{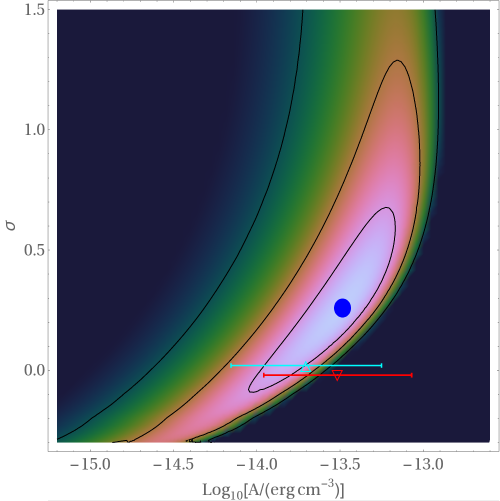}
                \includegraphics[height=0.33\textwidth]{Figures/ContR01th025pBst.png}
                \includegraphics[height=0.33\textwidth]{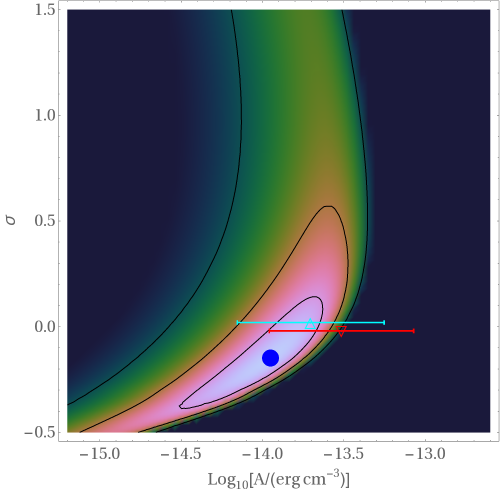}
            };
            \draw (-4.7, -2.0) node[text=white] {\scriptsize $p_{j=1..3}=1.84\pm0.64$};
            \draw (1.7, -2.0) node[text=white] {\scriptsize $p_{j=1..3}=1.84\pm0.64$};
            \draw (7.8, -2.0) node[text=white] {\scriptsize $p_{j=1..3}=1.84\pm0.64$};
       \end{tikzpicture}
    \end{center}
        \begin{center}
        \begin{tikzpicture}
            \draw (0, 0) node[inner sep=0]
            {
                \hspace{0.1cm}
                \includegraphics[height=0.33\textwidth]{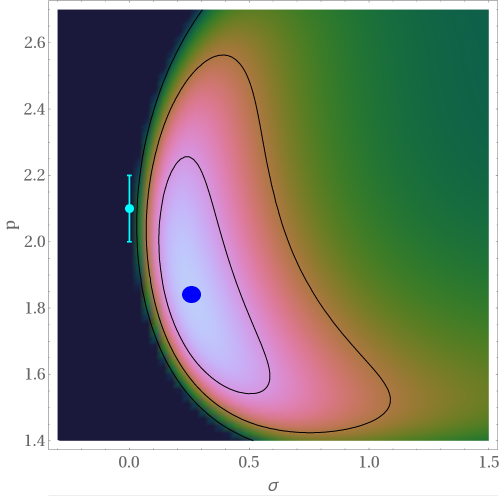}
                \includegraphics[height=0.33\textwidth]{Figures/ContR01th025ABst.png}
                \includegraphics[height=0.33\textwidth]{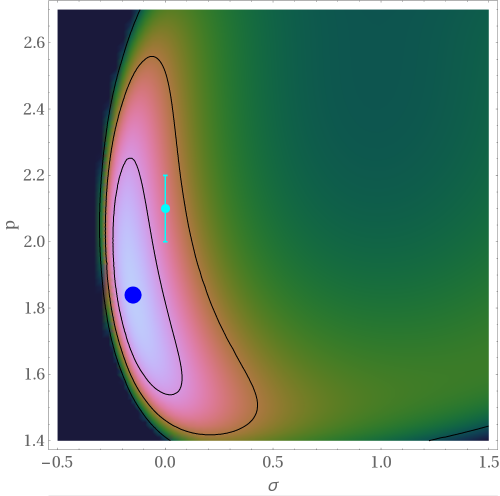}
            };
            \draw (-4.5, -2.0) node[text=white] {\scriptsize $A_{j=1..3}=10^{-13.5\pm0.8}$};
            \draw (1.7, -2.0) node[text=white] {\scriptsize $A_{j=1..3}=10^{-13.6\pm0.8}$};
            \draw (7.9, -2.0) node[text=white] {\scriptsize $A_{j=1..3}=10^{-14.0\pm0.8}$};
       \end{tikzpicture}
        \vspace{-0.6cm}
    \end{center}

   	\caption{
        Same as Fig.~\ref{fig:ModelContours} but varying the slope $\beta=2/3$ (left column), $0.768$ (nominal; middle column), and $1$ (right column).
    }
    \label{fig:ModelContoursBeta}
\end{figure*}

\begin{figure*}
    \begin{center}
        \begin{tikzpicture}
            \draw (0, 0) node[inner sep=0]
            {
                \includegraphics[height=0.33\textwidth]{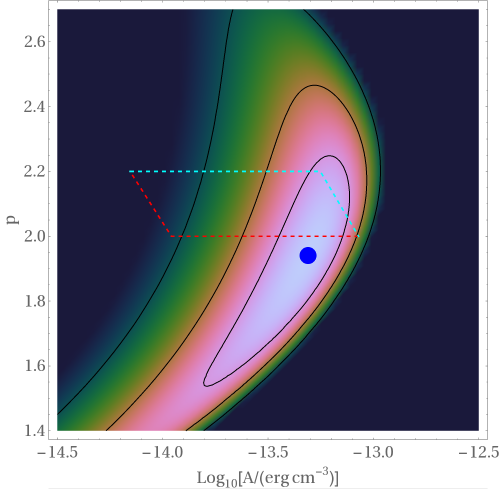}
                \includegraphics[height=0.33\textwidth]{Figures/ContR01th025SigBst.png}
                \includegraphics[height=0.33\textwidth]{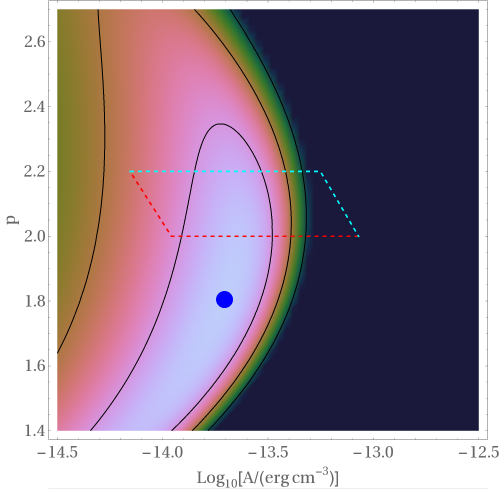}
            };
            \draw (-4.7, -2.0) node[text=white] {\scriptsize $\sigma_{j=1..4}=0.64\pm0.81$};
            \draw (1.7, -2.0) node[text=white] {\scriptsize $\sigma_{j=1..3}=0.10\pm0.48$};
            \draw (7.9, -2.0) node[text=white] {\scriptsize $\sigma_{j=1..3}=0.06\pm0.67$};
       \end{tikzpicture}
    \end{center}
    \begin{center}
        \begin{tikzpicture}
            \draw (0, 0) node[inner sep=0]
            {
                \includegraphics[height=0.33\textwidth]{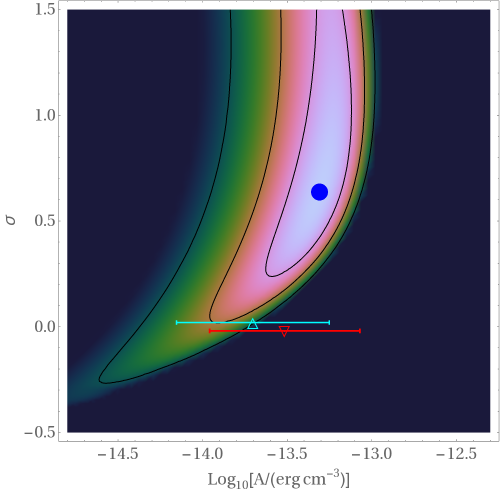}
                \includegraphics[height=0.33\textwidth]{Figures/ContR01th025pBst.png}
                \includegraphics[height=0.33\textwidth]{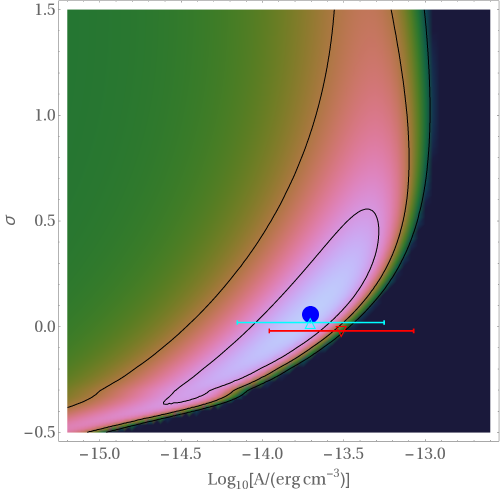}
            };
            \draw (-4.7, -2.0) node[text=white] {\scriptsize $p_{j=1..4}=1.94\pm0.43$};
            \draw (1.7, -2.0) node[text=white] {\scriptsize $p_{j=1..3}=1.84\pm0.64$};
            \draw (7.9, -2.0) node[text=white] {\scriptsize $p_{j=1..3}=1.80\pm0.80$};
       \end{tikzpicture}
    \end{center}
    \begin{center}
        \begin{tikzpicture}
            \draw (0, 0) node[inner sep=0]
            {
                \includegraphics[height=0.33\textwidth]{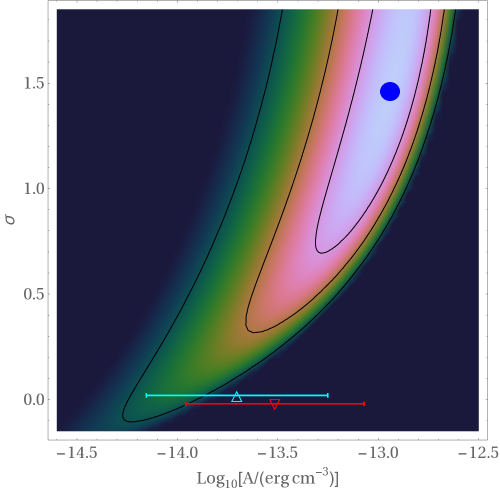}
                \includegraphics[height=0.33\textwidth]{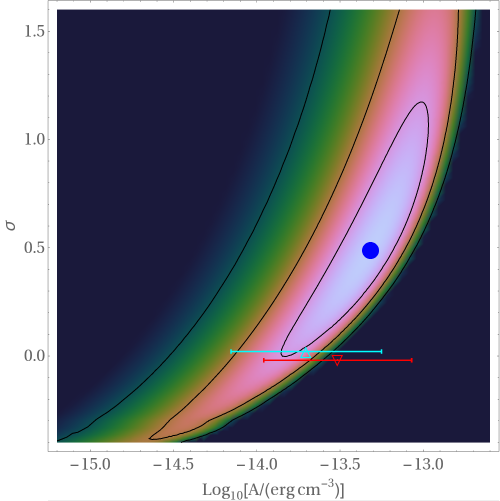}
                \includegraphics[height=0.33\textwidth]{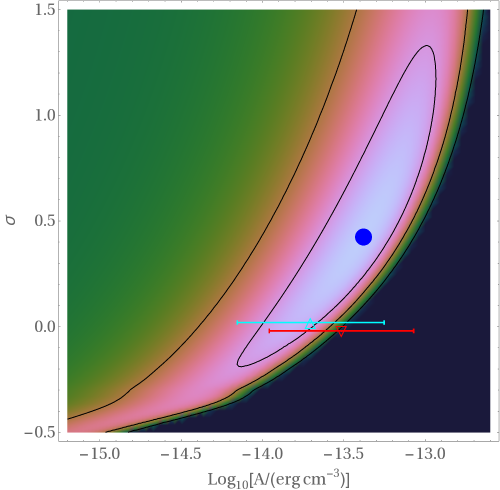}
            };
            \draw (-7.6, 2.6) node[text=white] {\scriptsize Assuming large,};
            \draw (-7.7, 2.3) node[text=white] {\scriptsize $\tau_c=0.3$ cores};
            \draw (-4.7, -1.6) node[text=white] {\scriptsize $p_{j=1..4}=1.93\pm0.42$};
            \draw (-1.3, 2.6) node[text=white] {\scriptsize Assuming large,};
            \draw (-1.4, 2.3) node[text=white] {\scriptsize $\tau_c=0.3$ cores};
            \draw (1.7, -2.0) node[text=white] {\scriptsize $p_{j=1..3}=1.82\pm0.63$};
            \draw (5, 2.6) node[text=white] {\scriptsize Assuming large,};
            \draw (4.9, 2.3) node[text=white] {\scriptsize $\tau_c=0.3$ cores};
            \draw (7.9, -2.0) node[text=white] {\scriptsize $p_{j=1..3}=1.78\pm0.80$};
       \end{tikzpicture}
    \end{center}
   	\caption{
        Same as Fig.~\ref{fig:ModelContours} (top two rows) but for different samples: nominal ($\theta_{500}>0\dgrdot2$; left column), non-compact ($\theta_{500}>0\dgrdot25$; middle), and extended ($\theta_{500}>0\dgrdot3$; right).
         Bottom row: same as middle row but for $\tau_c=0.3$.
    }
    \label{ModelContoursAll}
\end{figure*}

We repeat the analysis for different variants of the nominal hadronic model, finding that the results are robust and not particularly sensitive to any of the model or analysis parameters.
In particular, the following figures demonstrate the results when varying the parameters $\alpha$ (Fig.~\ref{fig:ModelContoursAlpha}), $\beta$ (Fig.~\ref{fig:ModelContoursBeta}), or $\tau_c$ (Fig.~\ref{ModelContoursAll}) of the nominal AB model \eqref{eq:nMCXC}, or the $\theta_{500}$ cut.

\clearpage
\onecolumn

\section{Cluster sample}
\label{sec:Sample}

The nominal 75-cluster sample in provided in Table ~\ref{tab:clusters}.

\vspace{+0.3cm}

{\small
	\begin{longtable}{llcccccccccl}
		 \caption{
            {
            \label{tab:clusters}
			 The nominal sample of 75 clusters, sorted by mass, with $\beta$-model parameters, where available. \\
            \textbf{Columns:} (1) Cluster catalog Name; (2) Alternate cluster name; (3) Galactic latitude (deg); (4) Galactic longitude (deg); (5) Mass within $\myR$ ($10^{14}M_\odot$); (6) Radius enclosing an overdensity $\delta=500$ (kpc); (7) Angular scale of $\myR$ (deg); (8) Cluster temperature (keV); (9) The $\beta$-model $\beta$ parameter; (10) The $\beta$-model central electron density ($10^{-3}\cm^{-3}$); (11) The $\beta$-model core radius (kpc); (12) Reference for $\beta$-model parameters: F04 \citep{FukazawaEtAl04}, C07 \citep{ChenEtAl07}.}
			}\\
		\hline
		MCXC Name &  Alt. Name & $b$ & $l$ &  $M_{500}$ &    $\myR$  &  $\theta_{500}$ &  $T$  &  $\beta$ &  $n_e$ &  $r_c $ &  Ref. \\
		(1) & (2) & (3) & (4) & (5) & (6) & (7) & (8) & (9) & (10) & (11)
        & (12) \\
		\hline
		\endfirsthead
		\hline
		MCXC Name &  Alt. Name & $b$ & $l$ &  $M_{500}$ &    $\myR$  &  $\theta_{500}$ &  $T$  &  $\beta$ & $n_e$ & $r_c $ &  Ref. \\
		\hline
		\endhead
		MCXC J0113.9-3145 & S141A & -83.27 & 257.67 & 0.11 & 341 & 0.24 &    &   &    &   & \\	
		MCXC J0920.0+0102 & MKW 1S & 33.06 & 230.94 & 0.13 & 359 & 0.28 &    &   &    &   & \\
		MCXC J1257.1-1339 & RXC J1257.1-1339  & 49.19 & 305.06 & 0.18 & 399 & 0.36 &    &   &    &   & \\
		MCXC J0933.4+3403 & UGC 05088 & 47.24 & 191.05 & 0.21 & 413 & 0.21 &    &   &    &   & \\
		MCXC J1304.2-3030 & RXC J1304.2-3030 & 32.27 & 306.20 & 0.23 & 428 & 0.50 &    &   &    &   & \\
		MCXC J1723.3+5658 & NGC 6370 & 34.34 & 85.21 & 0.23 & 428 & 0.22 &    &   &    &   & \\
		MCXC J1736.3+6803 & ZW 1745.6+6703 & 32.00 & 98.27 & 0.24 & 435 & 0.24 & 1.37 & 0.38 & 4.39 & 13 & \Fukazawa\\
		MCXC J2347.4-0218 & HCG 97 & -60.83 & 88.50 & 0.24 & 438 & 0.27 &  1.2 & 0.4 & 8.46 & 7 & \Fukazawa \\
		MCXC J1334.3+3441 & NGC 5223 & 78.09 & 74.98 & 0.25 & 442 & 0.25 &    &   &    &   & \\
		MCXC J0919.8+3345 & RXC J0919.8+3345 & 44.40 & 191.08 & 0.26 & 446 & 0.27 &   &   &   &   & \\
		MCXC J1253.0-0912 & HCG 62 & 53.67 & 303.62 & 0.27 & 455 & 0.42 &  1.1 & 0.4 & 23.18 & 5 & \Fukazawa\\
		MCXC J2249.2-3727 & S1065 & -62.35 &  3.29 & 0.31 & 472 & 0.23 &    &   &    &   & \\
		MCXC J1755.8+6236 & RXC J1755.8+6236 & 30.22 & 91.82 & 0.31 & 474 & 0.25 & 1.78 & 0.38 & 1.99 & 35 & \Fukazawa \\
		MCXC J2214.8+1350 & RX J2214.7+1350 & -34.13 & 75.17 & 0.32 & 477 & 0.26 &    &   &    &   & \\
		MCXC J0838.1+2506 & CGCG120-014 & 33.73 & 199.58 & 0.33 & 483 & 0.23 &    &   &    &   & \\
		MCXC J0916.1+1736 &   & 39.68 & 211.99 & 0.33 & 483 & 0.23 &    &   &    &   & \\
		MCXC J1615.5+1927 & NGC 6098 & 42.81 & 34.97 & 0.36 & 499 & 0.22 &    &   &    &   & \\
		MCXC J2111.6-2308 & AM2108 & -40.49 & 24.69 & 0.37 & 500 & 0.21 &    &   &    &   & \\
		MCXC J0907.8+4936 & VV 196 & 42.12 & 169.27 & 0.40 & 513 & 0.20 &    &   &    &   & \\
		MCXC J0125.6-0124 & A0194 & -63.00 & 142.07 & 0.40 & 516 & 0.39 &  1.9 & 0.4 & 1.11 & 81 & \Fukazawa\\
		MCXC J1050.4-1250 & USGC S152 & 40.40 & 262.76 & 0.41 & 522 & 0.46 &    &   &   &  & \\
		MCXC J1206.6+2811 & MKW4S & 80.02 & 204.27 & 0.42 & 523 & 0.26 & 1.90 & 0.38 & 5.32 & 18  & \Fukazawa \\
		MCXC J0249.6-3111 & S0301 & -63.96 & 229.00 & 0.45 & 536 & 0.32 &    &   &    &   & \\
		MCXC J1109.7+2145 & A1177 & 66.28 & 220.43 & 0.46 & 540 & 0.24 &    &   &    &   & \\
		MCXC J0036.5+2544 & ZWCL193 & -37.01 & 118.75 & 0.49 & 549 & 0.22 &    &   &    &   & \\
		MCXC J2224.7-5632 & S1020 & -50.71 & 334.28 & 0.50 & 552 & 0.22 &    &   &    &   & \\
		MCXC J2151.3-5521 & RXC J2151.3-5521 & -47.12 & 339.16 & 0.51 & 556 & 0.20 &   &   &   &   &  \\
		MCXC J2315.7-0222 & NGC 7556 & -56.28 & 76.06 & 0.58 & 585 & 0.30 &    &   &    &   & \\
		MCXC J0228.1+2811 & RX J0228.2+2811 & -30.01 & 147.57 & 0.66 & 608 & 0.24 &    &   &    &   & \\
		MCXC J2107.2-2526 & A3744 & -40.14 & 21.44 & 0.68 & 612 & 0.23 &    &   &    &   & \\
		MCXC J0110.0-4555 & A2877 & -70.85 & 293.05 & 0.71 & 625 & 0.36 &    &   &    &   & \\
        MCXC J1440.6+0328 & MKW 8 & 54.79 & 355.49 & 0.74 & 632 & 0.33 &    &   &  &   & \\
		MCXC J0000.1+0816 & RXC J0000.1+0816 & -52.48 & 101.78 & 0.74 & 630 & 0.22 &   &   &   &   & \\
		MCXC J0113.0+1531 & A0160 & -47.03 & 130.60 & 0.74 & 631 & 0.20 &    &   &    &   & \\
		MCXC J0058.9+2657 & RX J0058.9+2657 & -35.89 & 124.99 & 0.82 & 651 & 0.20 &    &   &    &   & \\
		MCXC J1733.0+4345 & IC 1262 & 32.07 & 69.52 & 0.86 & 664 & 0.30 &    &   &    &   & \\
        MCXC J1715.3+5724 & RXC J1715.3+5724 & 35.4 & 85.8 & 0.87 & 668 & 0.33 &   &   &   &   & \\
		MCXC J2101.8-2802 & A3733 & -39.60 & 17.77 & 0.92 & 679 & 0.25 &    &   &    &   & \\
        MCXC J1121.7+0249 & RXC J1121.7+0249 & 57.57 & 257.64 & 0.92 & 677 & 0.20 &   &   &   &   & \\
		MCXC J1253.2-1522 & A1631 & 47.49 & 303.57 & 0.98 & 692 & 0.21 &  2.28 & 0.85  & 0.49  & 15 & \Fukazawa \\
		MCXC J0040.0+0649 & A76 & -55.94 & 117.86 & 0.99 & 695 & 0.25 &    &   &    &   &  \\
		MCXC J1407.4-2700 & A3581 & 32.86 & 323.14 & 1.08 & 719 & 0.43 &  1.7 & 0.5 & 40.60 & 9& \Fukazawa \\
		MCXC J0525.5-3135 & A3341 & -31.09 & 235.17 & 1.09 & 718 & 0.26 &      &     &      &    &  \\
		MCXC J0341.2+1524 & IIIZw54 & -30.79 & 172.18 & 1.13 & 728 & 0.33 & 2.16 & 0.887 & 2.42 & 198 & \Chen \\
		MCXC J0828.6+3025 & A0671 & 33.15 & 192.75 & 1.15 & 728 & 0.21 &      &     &      &    &  \\
		MCXC J2310.4+0734 & Pegasus II & -47.54 & 84.15 & 1.17 & 735 & 0.24 &      &     &      &    &  \\
		MCXC J0003.2-3555 & A2717 & -76.49 & 349.33 & 1.20 & 739 & 0.21 &      &     &      &    &  \\
		MCXC J2338.4+2700 & A2634 & -33.09 & 103.48 & 1.22 & 746 & 0.34 &  3.5 & 0.4 & 2.17 & 75& \Fukazawa \\
		MCXC J0115.2+0019 & A0168 & -61.95 & 135.65 & 1.25 & 749 & 0.24 &      &     &      &    &  \\
		MCXC J0108.8-1524 & A0151 & -77.60 & 142.84 & 1.28 & 753 & 0.20 &      &     &      &    &  \\
		MCXC J0025.5-3302 & S0041 & -81.85 & 344.77 & 1.29 & 756 & 0.22 &      &     &      &    &  \\
		MCXC J2104.9-5149 & RXC J2104.9-5149 & -41.38 & 346.39 & 1.32 & 763 & 0.22 &      &     &      &    &  \\
        MCXC J2317.1+1841 & A2572a & -38.8 & 93.86 & 1.34 & 767 & 0.26 &   &   &   &   & \\
		MCXC J2324.3+1439 & A2593 & -43.18 & 93.45 & 1.42 & 783 & 0.26 &      &     &      &    &  \\
		MCXC J1539.6+2147 & A2107 & 51.53 & 34.40 & 1.49 & 796 & 0.27 &  3.5 & 0.6 & 5.76 & 85&  \\
		MCXC J1255.5-3019 & A3530 & 32.53 & 303.99 & 1.56 & 804 & 0.21 &  3.7 & 0.4 & 4.93 & 37&  \\
		MCXC J1454.5+1838 & A1991 & 60.50 & 22.79 & 1.68 & 823 & 0.20 &      &     &      &    &  \\
		MCXC J0125.0+0841 & A193 & -53.27 & 136.92 & 1.76 & 839 & 0.24 &      &     &      &    &  \\
		MCXC J2227.8-3034 & A3880 & -58.51 & 18.00 & 1.79 & 841 & 0.21 &      &     &      &    &  \\
		MCXC J2235.6+0128 & A2457 & -46.59 & 68.63 & 1.82 & 846 & 0.20 &      &     &      &    &  \\
		MCXC J1017.3-1040 & A0970 & 36.86 & 253.05 & 1.85 & 850 & 0.21 &      &     &      &    &  \\
		MCXC J2344.9+0911 & A2657 & -50.26 & 96.72 & 1.88 & 859 & 0.30 &  3.9 & 0.5 & 6.34 & 66& \Fukazawa \\
		MCXC J1252.5-3116 & RBS 1175  & 31.60 & 303.22 & 1.89 & 858 & 0.23 &      &     &      &    &  \\
		MCXC J1359.2+2758 & A1831 & 74.95 & 40.07 & 1.98 & 869 & 0.20 &      &     &      &    & \\
		MCXC J2347.7-2808 & A4038 & -75.86 & 25.14 & 2.04 & 886 & 0.41 &  2.9 & 0.5 & 34.53 & 14& \Fukazawa \\
		MCXC J0330.0-5235 & A3128 & -51.12 & 264.80 & 2.08 & 883 & 0.20 &      &     &      &    &  \\
		MCXC J0351.1-8212 & S0405 & -32.49 & 296.42 & 2.19 & 899 & 0.21 &  4.2 & 0.7 & 1.45 & 314& \Fukazawa \\
		MCXC J1257.2-3022 & A3532 & 32.48 & 304.43 & 2.34 & 920 & 0.24 &  4.3 & 0.5 & 3.84 & 102& \Fukazawa \\
		MCXC J0338.6+0958 & 2A0335 & -35.05 & 176.26 & 3.45 & 1055 & 0.42 &  3.1 & 0.5 & 117.8 & 11& \Fukazawa \\
		MCXC J0342.8-5338 & A3158 & -48.93 & 265.05 & 3.65 & 1067 & 0.26 &  5.0 & 0.6 & 6.17 & 108& \Fukazawa \\
		MCXC J0257.8+1302 & A0399 & -39.46 & 164.32 & 4.25 & 1117 & 0.23 &  6.4 & 0.6 & 4.53 & 142& \Fukazawa \\
		MCXC J1703.8+7838 & A2256 & 31.76 & 111.01 & 4.25 & 1122 & 0.28 &  6.5 & 0.9 & 2.67 & 400& \Fukazawa \\
        MCXC J0041.8-0918 & RXC J0041.8-0918 & -72.03 & 115.23 & 5.32 & 1210 & 0.31 &   &   &   &   & \\
		MCXC J1348.8+2635 & A1795 & 77.18 & 33.82 & 5.53 & 1224 & 0.28 &  5.8 & 0.6 & 15.62 & 83& \Fukazawa \\
		MCXC J0258.9+1334 & A0401 & -38.87 & 164.18 & 5.85 & 1242 & 0.25 &  8.0 & 0.7 & 5.64 & 207& \Fukazawa \\
        \hline
    \end{longtable}
}

\end{document}